\documentclass[aps,showpacs,prb,twocolumn,superscriptaddress]{revtex4}
\usepackage{epsfig}
\usepackage{amsfonts}
\usepackage{amsmath}
\usepackage{mathrsfs, amssymb, amsfonts, amsmath, mathtools}
\usepackage{bm}
\usepackage{float}
\usepackage[caption=false]{subfig}
\usepackage[utf8]{inputenc}
\usepackage[colorlinks,bookmarks=true,citecolor=blue,linkcolor=blue,urlcolor=blue]{hyperref}
\usepackage[usenames,dvipsnames]{xcolor}

\newcommand{\beq}{\begin{equation}}
\newcommand{\eeq}{\end{equation}}
\newcommand{\bea}{\begin{align}}
\newcommand{\eea}{\end{align}}
\newcommand{\nn}{\nonumber}

\newcommand{\ua}{\uparrow}
\newcommand{\da}{\downarrow}

\begin{document}

\title{Evidence for Majorana bound states in transport properties of hybrid structures based on helical liquids}
\author{Fran\c{c}ois Cr\'epin}
\affiliation{Institute for Theoretical Physics and Astrophysics,
University of W\"urzburg, 97074 W\"urzburg, Germany}
\author{Bj\"orn Trauzettel}
\affiliation{Institute for Theoretical Physics and Astrophysics,
University of W\"urzburg, 97074 W\"urzburg, Germany}
\author{Fabrizio Dolcini}
\affiliation{Dipartimento di Scienza Applicata e Tecnologia, Politecnico di Torino, 10129, Torino, Italy}
\affiliation{CNR-SPIN, Monte S.Angelo - via Cinthia, I-80126 Napoli, Italy}

\date{\today}

\begin{abstract}
Majorana bound states can emerge as zero-energy modes at the edge of a 
two-dimensional topological insulator in proximity to an ordinary s-wave 
superconductor. The presence of an additional ferromagnetic domain close 
to the superconductor can lead to their localization. We consider both 
N-S and S-N-S junctions based on helical liquids and study their 
spectral properties for arbitrary ferromagnetic scatterers in the normal 
region. Thereby, we explicitly compute Andreev wave-functions at zero 
energy. We show under which conditions these states form localized 
Majorana bound states in N-S and S-N-S junctions. Interestingly, we can 
identify Majorana-specific signatures in the transport properties of N-S 
junctions and the Andreev bound levels of S-N-S junctions that are 
robust against external perturbations. We illustrate these findings with 
the example of a ferromagnetic double barrier (i.e. a quantum dot) close 
to the N-S boundaries.

\end{abstract}


\maketitle

\section{Introduction}

As main requirements for topological quantum computers, Majorana bound states (MBS) in one-dimensional topological superconductors  have been in the focus of recent research. Originally,   MBS where shown by Kitaev~\cite{Kitaev01} to arise as localized edge states in a simple model of a 1D, spinless, $p$-wave superconductor~\cite{Kitaev01}. Subsequently, several groups proposed a possible experimental realization based on semiconductor nanowires with strong spin-orbit coupling, in proximity to an $s$-wave superconductor, and in the presence of a Zeeman field~\cite{Oreg10b,Lutchyn10}. In order to probe MBS in transport experiments, hybrid structures, namely normal-metal-superconductor (N-S) and Josephson (S-N-S) junctions, have been realized with InAs nanowires that produced data compatible with Majorana physics~\cite{Kouwenhoven12, Rokhinson12, Heiblum12}. Two kinds of transport signatures are generally considered. Tunnel current measurements in an N-S junction should lead to a robust zero-bias peak, signaling the presence of a zero energy mode -- the MBS -- at the interface~\cite{Law09}, while a fractional Josephson effect -- a $4\pi$ periodic supercurrent mediated by localized MBS -- is expected in S-N-S junctions~\cite{Kitaev01}. Both results were reported in recent experiments~\cite{Kouwenhoven12, Rokhinson12, Heiblum12}, although the situation remains to date controversial~\cite{Aguado12, Bagrets12, Lee12,  Rainis13, Marcus13, Lee14}. Inspired by these experiments, a remarkable attention has been devoted by theorists to the nanowire realizations. An interesting question is the fate of the localized MBS once contact is made with a normal lead, either in the N-S or the S-N-S case. Both numerical~\cite{Chevallier12} and analytical~\cite{Klinovaja12} works showed that the Majorana states completely delocalize in the normal lead~\cite{Chevallier12, Klinovaja12}, and, in Josephson junctions, typically transform into Andreev states~\cite{Chevallier12}, for superconductor phase differences away from $\pi$. Such a delocalization is robust to the inclusion of Coulomb interactions~\cite{Fisher12}.

 Recent experiments~\cite{Knez12, Hart13} carried on quantum spin Hall (QSH) insulators in proximity to $s$-wave superconductors may turn the tide. Without need for fine-tuning, normal edge states of QSH insulators form a true helical liquid~\cite{Kane05, Kane05b, Zhang06c, Konig07, Knez11}. Futhermore, Dirac mass defects -- such as the boundary between a ferromagnetic and a superconducting domain -- can also host Majorana states~\cite{Kitaev09, Fu09b}. In order to formulate precise predictions for transport experiments in N-S and S-N-S junctions based on helical liquids at the edge of topological insulators, it is therefore crucial to have a deeper understanding of  the formation of bound-states. Specific situations have been investigated by some groups, such as Josephson junctions in the tunneling regime~\cite{Fu09b, Schaffer11, Beenakker12, Houzet13} or in the presence of isolated ferromagnetic impurities~\cite{Taddei13, Zhang13}, magneto-Josephson effects~\cite{Refael13}, as well as N-S junctions with a quantum dot and a small Zeeman field~\cite{Wimmer13}. However, a more general approach to the problem was missing so far. With this perspective in mind, we have derived a general result
for the N-S and the S-N-S junctions with helical liquids in the presence of an {\it arbitrary} ferromagnetic domain,  including, in particular, the case of two ferromagnetic barriers.
More specifically, for the N-S case we have obtained a general formula for the Andreev reflection probability, which shows that, in addition to the zero excitation energy modes that are always perfectly Andreev reflected,  many resonant, Fabry-P\'erot like, peaks can appear at non-zero energies,   related to virtual bound states at the ferromagnet-superconductor interface. As for the S-N-S case, we have determined the expression for the Andreev-bound levels, which shows that a zero-energy Andreev bound state at phase difference equal to $\pi$ is stable, independent of the shape and strength of the ferromagnetic domain. Explicit results are shown for a ferromagnetic quantum dot realized by two sharp ferromagnetic barriers. Furthermore, for the particular case of a single ferromagnetic barrier of finite length, we provide explicit expressions of the Majorana wave functions, localized on either side of the barrier, in both N-S and S-N-S configurations. 



{\it Outline and summary of results. ---} The paper is organized as follows. In section~\ref{sec:spectral}, we start by reviewing the spectral properties of the Bogoliubov-de Gennes theory for inhomogeneous superconductors, in the case of broken spin rotation invariance. We discuss in particular the construction of Majorana states. In section~\ref{sec:NS}, we investigate the transport properties of N-S junctions in the presence of a ferromagnetic scatterer and, in section~\ref{sec:SNS}, we discuss properties of the Andreev bound levels for S-N-S junctions. In section \ref{sec:conclu}, we discuss the implications of our findings for the detection of Majorana bound states in experiments.

\begin{figure}
        \centering
        \subfloat{\includegraphics[width=6cm,clip]{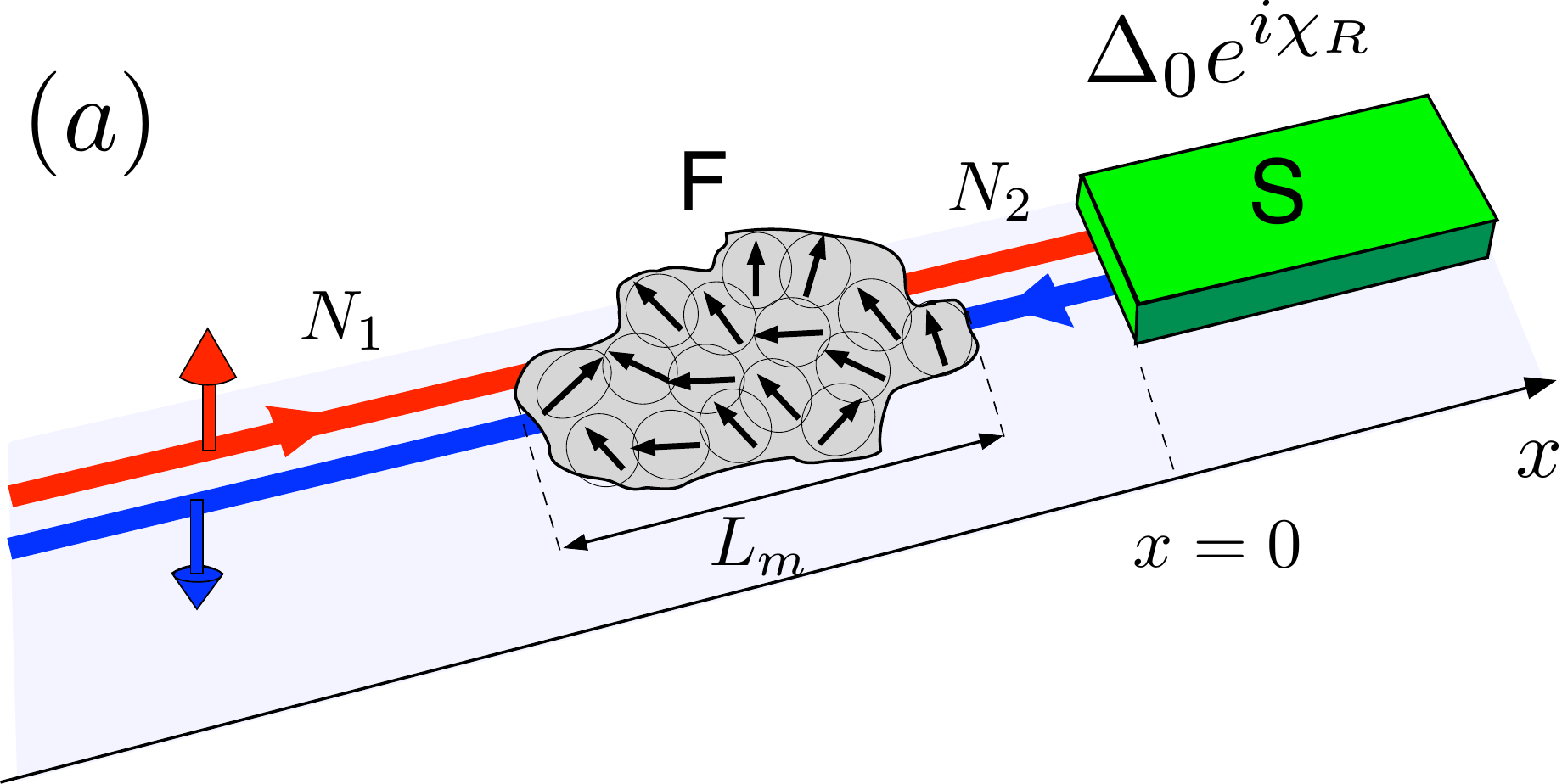}}
        
		\subfloat{\includegraphics[width=6cm,clip]{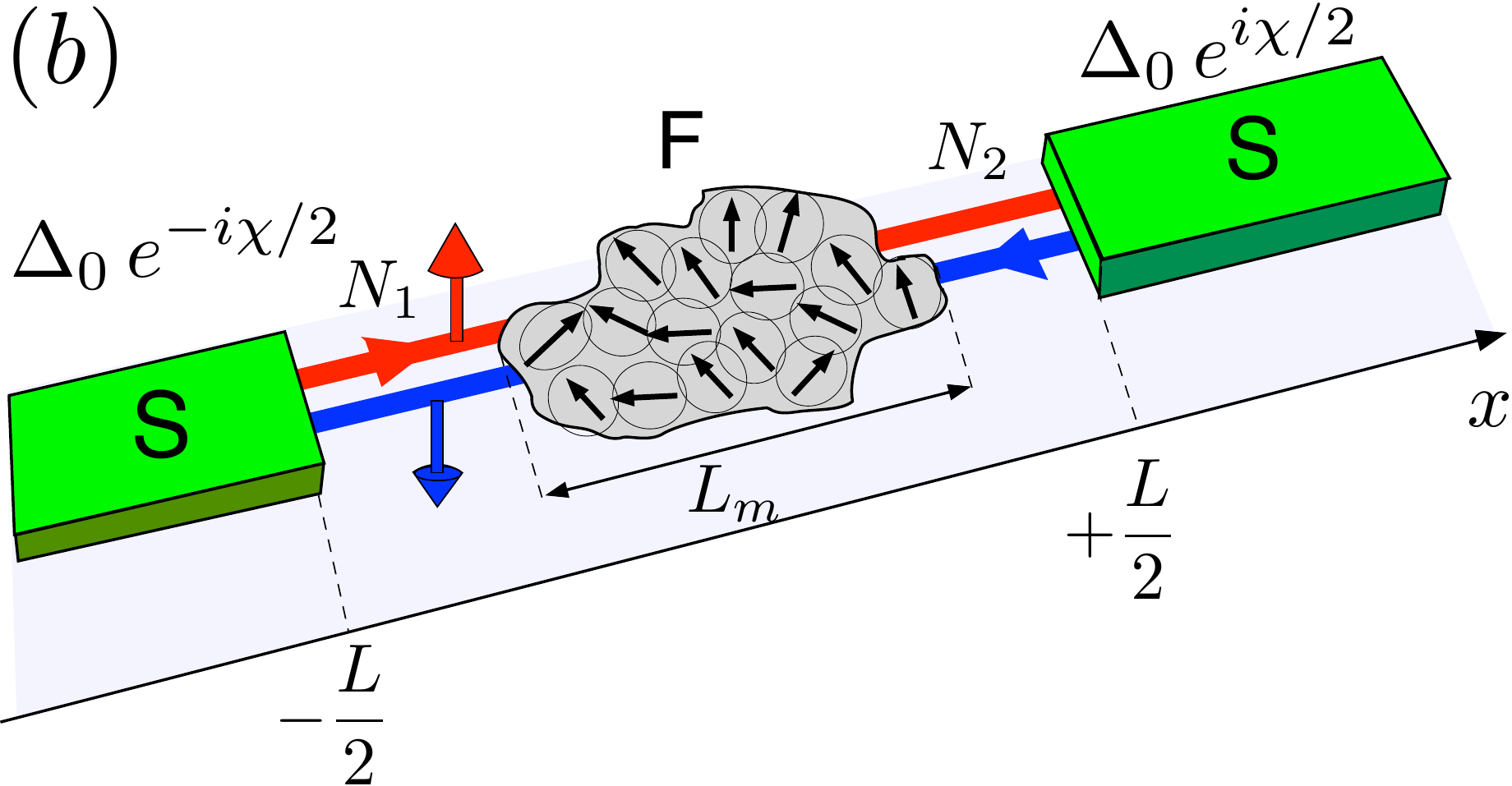}}
		 \caption{(Color online) N-S and S-N-S junctions realized with helical edge states, in the presence of a ferromagnetic domain F in the normal region. The regions $N_1$ and $N_2$ represent the interface regions where helical states propagate freely.  }\label{Fig:S-F-S-setup-Fig}
\end{figure}






\section{Spectral properties of the Bogoliubov-de Gennes theory}
\label{sec:spectral}

\subsection{Hamiltonian of the system}

We start by reviewing some general properties of the hybrid structures we shall consider thereafter.  A helical liquid consists of a pair of edge states in a quantum spin Hall insulator (QSHI), where  the group velocity is locked to the spin orientation. The helical liquid is contacted to one or two $s$-wave superconducting electrodes. Additionally, the presence of a ferromagnet along the edge induces an arbitrary Zeeman coupling in the normal region (see Fig.~\ref{Fig:S-F-S-setup-Fig}). The Hamiltonian of such a structure is given by
\beq
H = H_0 + H_Z + H_{\Delta}\;, \label{HTOT}
\eeq
where $H_0$ is the Hamiltonian of the helical liquid, $H_Z$ describes the Zeeman coupling and $ H_{\Delta}$ the proximity induced pairing potential. For definiteness, and without loss of generality,  we assume that for the edge states  the spin quantization axis is well-defined and points along the $z$ direction, and that right-(left-) moving electrons are characterized by spin-$\uparrow$ (spin-$\downarrow$),  so that
\begin{align}
H_0 =  \int dx \, \, (\psi^\dagger_{R\uparrow} , \psi^\dagger_{L\downarrow} ) \, \left[ v_F p_x \sigma_z -\mu \right] \, \left(\begin{array}{l}  \psi^{}_{R\uparrow}  \\   \psi^{}_{L\downarrow} \end{array} \right)\;,
\end{align}
where $p_x=-i \hbar \partial_x$ and $\mu$ denotes the chemical potential. The presence of the ferromagnetic domain is accounted for by the term 
\beq
H_Z=\int dx \, \, (\psi^\dagger_{R\uparrow} , \psi^\dagger_{L\downarrow} ) \,  \mathbf{m}(x)\cdot \boldsymbol{\sigma} \, \left(\begin{array}{l}  \psi^{}_{R\uparrow}  \\   \psi^{}_{L\downarrow} \end{array} \right)\;,
\eeq
where $\boldsymbol{\sigma} = (\sigma_x,\sigma_y,\sigma_z)$ are the Pauli matrices and 
$\mathbf{m}(x) = (m_\parallel \cos \phi, m_\parallel \sin \phi, m_z)$ is the space-dependent magnetization vector with $m_\parallel(x) >0$. Here, $\phi(x)$  denotes the local absolute value and angle of the in-plane magnetization, respectively, and $m_z$ the perpendicular magnetization component. The electron field operator $\psi^\dagger_{R\uparrow}(x)$ (resp. $\psi^{}_{R\uparrow}(x)$) creates (resp. annihilates) a right mover with spin $\ua$, while $\psi^\dagger_{L\downarrow}(x)$ (resp. $\psi^{}_{L\downarrow}(x)$) creates (resp. annihilates) a left mover with spin $\da$. Finally, the superconducting pairing potential is given by
\beq
H_{\Delta} =  \int dx \,\left[ \Delta(x) \, \psi^\dagger_{R\uparrow}   \psi^\dagger_{L\downarrow} + \Delta^*(x) \psi^{}_{L\downarrow} \psi^{}_{R\uparrow} \right]\;.
\eeq
The Hamiltonian (\ref{HTOT}) can be rewritten in a Bogoliubov-de Gennes (BdG) form 
\begin{equation}
H=\frac{1}{2} \int dx \; \Psi^\dagger \mathcal{H}_{\rm BdG} \Psi^{} \label{H-in-BdG form}
\end{equation}
by introducing the Nambu spinor $\Psi^\dagger= \, (\psi^\dagger_{R\uparrow} , \psi^\dagger_{L\downarrow} ,  \psi^{}_{L\downarrow} , -\psi^{}_{R\uparrow} )$ and the Hamiltonian matrix
\begin{equation}
\mathcal{H}_{\rm BdG}= \begin{pmatrix}
 H^{e}_0   &   \Delta \sigma_0       \\  
 & \\
 \Delta^* \sigma_0   &  H_0^{h} \end{pmatrix} \quad.\label{H-BdG} 
\end{equation}
In Eq.~(\ref{H-BdG}), 
\begin{subequations}
\begin{align}
H_0^{e} &= v_F \sigma_z p_x -\mu \sigma_0 + \mathbf{m}(x)\cdot \boldsymbol{\sigma}\;, \\
H_0^{h} &=-\mathcal{T} H_0^{e} \mathcal{T}^{-1} = -\sigma_y (H_0^{e})^* \sigma_y= \nonumber \\
& =-v_F \sigma_z p_x +\mu \sigma_0 + \mathbf{m}(x)\cdot \boldsymbol{\sigma}\;.
\end{align}
\end{subequations}
are the particle and hole sector diagonal blocks, respectively,   $\sigma_0$ denotes the identity matrix in spin space~\footnote{Everywhere in the paper $\sigma_x,\sigma_y,\sigma_z$ are Pauli matrices acting on spin degrees of freedom while $\tau_x,\tau_y,\tau_z$ are Pauli matrices acting on Nambu (particle-hole) degrees of freedom.}, and $\mathcal{T} = \mathcal{K} i\sigma_y $ is the time-reversal operator, with $\mathcal{K}$ the complex conjugation.  \\

For the moment, we keep an arbitrary profile for both the pairing potential $\Delta(x)$ and the ferromagnetic coupling $\mathbf{m}(x)$. Later, we shall specify $\Delta(x)$ for the case of N-S and S-N-S junctions, whereas general results will be given for an arbitrary profile $\mathbf{m}(x)$.

\subsection{Quasi-particle states}

The Hamiltonian (\ref{H-in-BdG form}) can be written in a diagonal form,
\beq
H = \sum_{\varepsilon_n \geq 0} \sum_{j} \varepsilon_n \, \gamma^\dagger_{\varepsilon_n,j}\gamma^{}_{\varepsilon_n,j}\;, \label{Eq:H_qp}
\eeq
where $\gamma^\dagger_{\varepsilon_n,j }$ and $\gamma^{}_{\varepsilon_n, j}$ respectively create and annihilate a fermionic quasi-particle with positive excitation energy~$\varepsilon_n$, with respect to a ground state, whose energy has been set to zero. The label $j$ accounts for possible degeneracies, examples of which will be given in later sections. 
The diagonalization (\ref{Eq:H_qp}) is achieved from (\ref{H-in-BdG form}) through the following ansatz
\begin{equation}
\Psi(x)  =  \! \sum_{\varepsilon_n \ge 0} \sum_j  \left( \varphi_{\varepsilon_n,j}(x) \gamma^{}_{\varepsilon_n, j} + \left[\mathcal{C}\varphi_{\varepsilon_n,j}\right](x) \gamma^\dagger_{\varepsilon_n, j} \right)  \label{Eq:rotation}  ,
\end{equation}
where
\begin{equation}
\varphi_{\varepsilon_n,j} \doteq ( u_{\varepsilon_n,j,\ua} , u_{\varepsilon_n,j,\da} , v_{\varepsilon_n,j,\da}, v_{\varepsilon_n,j,\ua})^T \label{wavefunction}
\end{equation}
is a solution of the BdG equation~\cite{deGennesBook}
\beq
\mathcal{H}_{\textrm{BdG}} \, \varphi_{\varepsilon_n,j}  =  \varepsilon_n \varphi_{\varepsilon_n,j} \;, \label{eq:HBdG}
\eeq
and $\mathcal{C}\varphi_{\varepsilon_n,j}$ is its charge-conjugated wavefunction. 
Here, we have introduced the anti-unitary charge-conjugation operator $\mathcal{C} = \mathcal{K} U_C$, with $U_C = \tau_y \otimes \sigma_y $, and $\mathcal{K}$ the complex conjugation. The relations~\eqref{Eq:rotation} can be inverted, and the quasi-particle operator $\gamma_{\varepsilon_n,j}$ expressed as 
\begin{eqnarray}
\lefteqn{\gamma_{\varepsilon_n,j} = \int dx \, (\varphi_{\varepsilon_n,j}^*(x))^T \, \Psi(x) = } & & \label{gamma} \\
&=&\int dx\;    \Big[ u_{\varepsilon_n,j, \ua}^*(x) \psi_{R\ua}(x) + u_{\varepsilon_n,j,\da}^*(x) \psi_{L\da}(x)  \nn \\
& &\left.  \hspace{1cm}+  v_{\varepsilon_n,j,\da}^*(x) \psi^\dagger_{L\da}(x)   -v_{\varepsilon_n,j,\ua}^*(x)\psi^\dagger_{R\ua}(x) \right]\;, \nonumber 
\end{eqnarray}
\\As can be seen from Eq.~(\ref{H-BdG}), in Eq.~(\ref{eq:HBdG})  the superconducting pairing potential couples a right-(left-) moving electron $u_{\varepsilon_n,j,\ua}$ ($u_{\varepsilon_n,j,\da}$) to a left-(right) moving hole $v_{\varepsilon_n,j,\da}$, ($v_{\varepsilon_n,j,\ua}$). Furthermore, while the $m_z$ component of the magnetization of the ferromagnetic domain preserves spin as a good quantum number, the in-plane magnetization $m_\parallel$ couples the dynamics of right and left moving electrons (holes), $u_{\varepsilon_n,j,\ua}$ ($v_{\varepsilon_n,j,\ua}$) and $u_{\varepsilon_n,j,\da}$ ($v_{\varepsilon_n,j,\da}$). Thus, differently from the standard treatment of FS junctions, where only the $m_z$ magnetization is considered  (see, e.g. Refs.~\onlinecite{DeJong95, Taddei05, Buzdin05}), no decoupling of the BdG equations into two $2\times 2$ independent blocks occurs here, and the solutions of the BdG equations are always four-component wave functions $\varphi_{\varepsilon_n,j}$. The problem is therefore closer to that of hybrid structures with spin-orbit coupling, with the important difference that the ferromagnetic coupling does break time reversal symmetry~\cite{Yokoyama05, Dimitrova06}.

\subsection{Particle-hole symmetry and Majorana states}
It is well-known that the BdG Hamiltonian (\ref{H-BdG}) exhibits a built-in particle-hole symmetry. Indeed, one has   $\mathcal{C} \mathcal{H}_{\textrm{BdG}} \mathcal{C}^{-1}= U_C^\dagger \mathcal{H}_{\textrm{BdG}}^* U_C = - \mathcal{H}_{\textrm{BdG}}$.  
The particle-hole symmetry entails that, if  $\varphi_{\varepsilon_n,j}$ is an eigenstate of $\mathcal{H}_{\textrm{BdG}}$  with energy $ \varepsilon_n$, then the charge conjugated state, $\mathcal{C} \varphi_{\varepsilon_n,j}$, is  an eigenstate of $\mathcal{H}_{\textrm{BdG}}$  with energy $-\varepsilon_n$. Introducing the following notation, $\mathcal{C} \varphi_{\varepsilon_n,j} = \varphi_{-\varepsilon_n,j^c}$, the relation between components of charge-conjugated states reads
\begin{eqnarray}
\varphi_{-\varepsilon_n,j^c}  &= &(u_{- \varepsilon_n,j^c,\ua},u_{-\varepsilon_n,j^c,\da},v_{-\varepsilon_n,j^c,\da},v_{-\varepsilon_n,j^c,\ua}) \label{Eq:charge-conj}  \\
&=&\mathcal{C} \varphi_{\varepsilon_n,j} = (-v_{\varepsilon_n,j,\ua}^*,v_{\varepsilon_n,j,\da}^*,u_{\varepsilon_n,j,\da}^*,-u_{\varepsilon_n,j,\ua}^*) \nonumber \;. 
\end{eqnarray}
and, combining Eqs.~\eqref{Eq:charge-conj} and (\ref{gamma}), one obtains
\beq
\gamma_{\varepsilon_n,j}^\dagger   = \gamma_{-\varepsilon_n,j^c} \;. \label{Eq:gamma_gamma_dagger}
\eeq
The latter equation has two important consequences. First, it allows for the potential existence of Majorana fermions, quasi-particles that are equal to their antiparticles ($\gamma^\dagger=\gamma^{}$). Indeed, from Eq.~(\ref{Eq:gamma_gamma_dagger}) a quasi-particle excitation is a Majorana fermion  if -- and only if -- it fulfills {\it two} conditions, namely it (i) has vanishing energy and (ii) is invariant under charge-conjugation, $j^c=j$. These conditions amount to state that a Majorana wavefunction $\varphi(x)$ is a kernel solution of the BdG equations, $\mathcal{H}_{\textrm{BdG}}\, \varphi=0$, that fulfills the constraint $\mathcal{C}\varphi = \varphi$, that is,
\begin{equation}
\left\{ \begin{array}{lcl} 
u^{}_{\ua} &=& -v^*_{\ua} \\
u^{}_{\da} &=& v^*_{\da} 
\end{array}\right. \quad.
\end{equation}
Any zero-energy fermionic quasi-particle $\gamma_{0,j}$ that is not Majorana-like ($j \neq j^c$) can always be decomposed as $\gamma_{0,j}=c_{+}+i c_{-}$, where $c_{+} \doteq \gamma_{0,j}+\gamma_{0,j^c}$ and $c^{}_{-}\doteq-i\gamma_{0,j}+i\gamma_{0,j^c}$ are two Majorana fermions ($c^{}_\pm=c^\dagger_\pm$). Because $c_\pm$ are linear combinations of quasi-particles within the same energy subspace, they are also proper excitations of the system. The corresponding Majorana wave-functions are $\varphi_{+} = \varphi_{0,j} +   \varphi_{0,j^c}$ and $\varphi_{-} = -i \varphi_{0,j} + i\varphi_{0,j^c}$. Being zero-energy states, they are likely to be bound in regions of space where the superconducting gap closes, and, in certain situations, even spatially separated. In the next sections, we clarify the conditions of emergence of such Majorana {\it bound-states} in hybrid structures based on helical liquids, and provide their explicit expressions in some relevant cases. We also notice that, formally, Majorana fermions can be constructed out of any pair of charge-conjugated fermionic quasi-particles $\gamma_{\varepsilon_n,j}$ and $\gamma_{-\varepsilon_n,j^c}$, of finite energy $\varepsilon_n>0$. Indeed, taking two complex numbers $\alpha^{}_{+}$ and $\alpha^{}_{-}$ such that $\alpha^{}_{+} \alpha_{-}^* - \alpha^*_{+} \alpha^{}_{-} \neq 0$, one can define two linearly independent, although not necessarily orthogonal, Majorana operators $c_{\pm} = \alpha_\pm \gamma_{\varepsilon_n,j} + \alpha^*_\pm \gamma_{-\varepsilon_n,j^c}$. 
Again, due to Eq.~(\ref{Eq:gamma_gamma_dagger}) one has $c^{}_\pm=c^\dagger_\pm$. However, for $\varepsilon_n \neq 0$,  such Majorana particles are not proper excitations of the system, as they are built up out of quasi-particles with opposite energies. The related wave functions are not stationary states of the BdG equations. A comprehensive discussion of the Majorana nature of Bogoliubov particles in superconductors is given by Chamon {\it et al.} in Ref.~\onlinecite{Chamon10}.\\

The second implication of Eq.~(\ref{Eq:gamma_gamma_dagger}) is concerned with the physical interpretation of the diagonalized Hamiltonian (\ref{Eq:H_qp}). Indeed in Eq.~(\ref{Eq:H_qp}) the ground-state $|0\rangle$ was taken to be the vacuum annihilated by all quasi-particles $\gamma_{\varepsilon_n,j}$ with energies $\varepsilon_n >0$, that is, $\gamma_{\varepsilon_n,j}|0\rangle=0$. Eq.~\eqref{Eq:gamma_gamma_dagger} indicates that the ground-state can equivalently be regarded as the filled sea of quasi-particles with energies $\varepsilon_n <0$. Furthermore, the Hamiltonian (\ref{Eq:H_qp}) can be given other equivalent and somewhat more symmetric expressions, which turn out to be particularly useful in situations where fermion number parity plays a role~\cite{Fu09b, Beenakker12, Crepin13b}. In particular, by rewriting
\beq
H = \sum_{\varepsilon_n \geq 0 \,, j}  \varepsilon_n \left( \gamma^\dagger_{\varepsilon_n,j }\gamma^{}_{\varepsilon_n,j } - \frac{1}{2}  \right) +   \frac{1}{2}\sum_{\varepsilon_n \geq 0\, ,j} \varepsilon_n\;. \label{H-equiv-1}
\eeq
the system is --up to a shift of the ground-state energy-- adequately described as a collection of single fermionic levels that can either be occupied, with an energy of $\varepsilon_n/2$, or empty, with an energy of $-\varepsilon_n/2$. In this language, the ground-state is characterised by all empty levels. The form \eqref{H-equiv-1}   is useful, for instance,  in the study of Josephson junctions, where  quasi-particles correspond  to Andreev bound states. Indeed it  allows for a clear and simple description of  the interplay between superconductivity and helicity in terms of a change in fermion parity when the phase difference across the junction is changed by $2\pi$~\cite{Fu09b, Crepin13b}.  Similarly, one can also write 
\beq
H = \frac{1}{2} \sum_{\varepsilon_n} \sum_{j}\varepsilon_n   \Gamma^\dagger_{\varepsilon_n,j }\Gamma^{}_{\varepsilon_n,j }   +  \frac{1}{2} \sum_{\varepsilon_n \geq 0\, ,j } \varepsilon_n\;,  \label{H-equiv-2}
\eeq
where $\Gamma_{\varepsilon_n,j }=\gamma_{\varepsilon_n,j }$ for $\varepsilon_n>0$ and $\Gamma_{\varepsilon_n,j }=\gamma_{\varepsilon_n,j^c}$ for $\varepsilon_n<0$.\\\\

\section{N-S junctions}
\label{sec:NS}


\subsection{A condition for perfect Andreev reflection}
\label{sec:NS-a}
We start with considering the case of an interface between the helical state and one superconductor, as   depicted in Fig.~\ref{Fig:S-F-S-setup-Fig}(a). Helicity forbids normal scattering at the NS interface and, as a consequence, in the normal region $N_2$ an electron (resp. a hole) is perfectly Andreev reflected as a hole (resp. an electron) at any subgap excitation energy $\varepsilon < \Delta_0$~\cite{Adroguer10}. On the other hand, a ferromagnetic~(F) region, as shown in Fig.~\ref{Fig:S-F-S-setup-Fig}, can induce normal backscattering. Let us  first focus on this effect:  An arbitrary ferromagnetic domain can be described in terms of a 2$\times$2 unitary scattering matrix, which  can be written in the following polar representation (all symmetries are broken)
\begin{widetext}
\begin{equation}
S_0^{e}(\varepsilon)= \,\left( 
\begin{array}{lcl} 
r_e(\varepsilon)  &  t^\prime_e(\varepsilon)  \\ & \\
t_e(\varepsilon)  & r^\prime_e(\varepsilon) 
\end{array} \right) =  e^{i\Gamma_m(\varepsilon)} \,   \left( 
\begin{array}{lcl} 
-i  \,e^{+i\Phi_m(\varepsilon)}   \,   \sqrt{1-T_\varepsilon}  &    \,e^{i \chi_m(\varepsilon)} \,  \sqrt{T_\varepsilon}  \\ & \\
   \, e^{-i \chi_m(\varepsilon)} \,  \sqrt{T_\varepsilon} &-i  \,e^{-i\Phi_m(\varepsilon)}   \,   \sqrt{1-T_\varepsilon} 
\end{array} \right)\;, \label{Eq:S0-p-gen}
\end{equation}
\end{widetext}
where $T_\varepsilon = |t_e(\varepsilon)|^2$ is the transmission coefficient of the F domain at excitation energy $\varepsilon$. Based on specific realizations, some of which will be further discussed below, one can ascribe a physical meaning to the other parameters as well. Indeed, $\Gamma_m \sim k_F L_m$ and $\Phi_m \sim k_F x_0$ (with $k_F=\mu/\hbar v_F$) are dynamical phases related to  the spatial extension $L_m$ of the ferromagnetic domain, and to the location $x_0$ of  its center with respect to the origin, respectively, whereas $\chi_m \sim m_z L_m$ is the relative phase shift between spin-$\uparrow$ and spin-$\downarrow$ electrons,  accumulated along the domain due to the Zeeman coupling in $z$ direction. As far as the F domain is concerned, scattering of electrons and holes are decoupled. The scattering matrix for holes is easily obtained from Eq.~\eqref{Eq:S0-p-gen}, by noticing that if $u(\varepsilon)$ is a solution of $H_0^{e} u(\varepsilon) = \varepsilon u(\varepsilon)$, in the electron sector, then $v(\varepsilon) = i\sigma_y u(-\varepsilon)$ is a solution $H_0^{h} v(\varepsilon) = \varepsilon v(\varepsilon)$, in the hole sector. From this one can deduce the important relation~\cite{Badiane13}
\beq
 S_0^{h}(\varepsilon)=-\sigma_z \, {S_0^{e}}^*(-\varepsilon)\, \sigma_z \;. \label{Eq:scatt_e_h}
\eeq
The whole scattering matrix  $S_N(\varepsilon) = \textrm{Diag}[S_0^{e}(\varepsilon),S_0^{h}(\varepsilon)] $ relates the scattering amplitudes $b$'s, for electrons and holes  out-going from F in regions $N_1$ and $N_2$, to the in-coming scattering amplitudes $a$'s, 
\begin{align}
(b_{e,1},b_{e,2},b_{h,1},b_{h,2})^T = S_N(\varepsilon)(a_{e,1},a_{e,2},a_{h,1},a_{h,2})^T\;. \label{Eq:scatt_amplitudes}
\end{align}
One has to combine such normal scattering with the Andreev scattering  at the interface, which couples electrons and holes. For sub-gap excitations energies, perfect Andreev reflection at the NS interface relates electron and hole scattering amplitudes in region $N_2$ through
\begin{eqnarray}
\left( \begin{array}{c}  a_{e,2} \\ \\  a_{h,2} \end{array} \right) =   \alpha(E) \! \!\left( \begin{array}{cc}     0 &   e^{i \chi_R } \! \\ &   \!\\      e^{-i \chi_R  } & 0 \end{array} \right)  \! \left( \begin{array}{c}  b_{e,2}   \\ \\  b_{h,2}\end{array} \right)  \label{Andreev-NS}
\end{eqnarray}
with $\alpha(\varepsilon) = \exp [-i \textrm{arccos} (\varepsilon/\Delta_0)]$. Here we have assumed that the origin of the $x$-axis is at the interface, as customary for the case of a N-S junction (for the S-N-S case there are two interfaces and a different choice is more suitable, as we shall see).  Simple algebra leads to the reflection matrix,  relating electron and hole amplitudes in region $N_1$ as
\beq
\left( \begin{array}{c}
b_{e,1} \\ \\ b_{h,1} \end{array} \right) 
= \left( \begin{array}{cc} r_{ee} & r_{eh} \\ & \\ r_{he} & r_{hh} \end{array} \right) \left( \begin{array}{c} \, a_{e,1}\\ \\ a_{h,1} \end{array}\right)  \label{r-matrix-NS}
\eeq
with
\begin{eqnarray}
r_{ee} &= \displaystyle\frac{r_e-\alpha^2 r^\prime_h \, {\rm det} S^{e}_0}{1-\alpha^2 r^\prime_e r^\prime_h}\;,\;  r_{eh} = \frac{\alpha \, t^\prime_e t_h \, e^{i \chi_R}}{1-\alpha^2 r^\prime_e r^\prime_h}\;, \nn \\
r_{he} &= \displaystyle \frac{\alpha \, t_e t^\prime_h \, e^{-i \chi_R}}{1-\alpha^2 r^\prime_e r^\prime_h}\;,\;   r_{hh} = \frac{r_h-\alpha^2 r^\prime_e \, {\rm det} S^{h}_0}{1-\alpha^2 r^\prime_e r^\prime_h}\;. 
\end{eqnarray}
\\It follows that, $|r_{ee}|^2=|r_{hh}|^2 \equiv R_N$, 
$|r_{eh}|^2=|r_{he}|^2 \equiv R_A$ and $R_N+R_A =1$, the latter relation reflecting current conservation.  Notice that Eq.(\ref{r-matrix-NS}) entails that, for the N-S junction, for each eigenvalue $\varepsilon$ of the BdG equation, there are two degenerate states ($j=1,2$ in Eq.(\ref{Eq:H_qp})), corresponding to the injection  of an electron and the injection of a hole, respectively, from  region $N_1$ (see Fig.\ref{Fig:S-F-S-setup-Fig}(a)).  Using Eqs.~\eqref{Eq:S0-p-gen} and \eqref{Eq:scatt_e_h} we arrive at the following general expression for the Andreev reflection probability at an NS interface in a helical liquid:
\begin{widetext}
\beq
R_A(\varepsilon) = \frac{T_\varepsilon \, T_{-\varepsilon}}{(1-\sqrt{(R_\varepsilon R_{-\varepsilon}})^2+4 \cos^2\left[ \arccos\frac{\varepsilon}{\Delta_0} +\Phi_m^A(\varepsilon)\right] \sqrt{R_\varepsilon R_{-\varepsilon}}}\;, \label{Eq:RA-fin-gen}
\eeq
\end{widetext}
where  $R_\varepsilon = 1-T_\varepsilon$ is the reflection coefficient of the F domain and $\Phi_m^A(\varepsilon) = \big(\Phi_m(\varepsilon)-\Phi_m(-\varepsilon)\big)/2$ is an odd function of the energy $\varepsilon$ that is extracted from the scattering matrix (\ref{Eq:S0-p-gen}). One immediately sees from Eq.~\eqref{Eq:RA-fin-gen} that, independently of the parameters $\Phi_m^A(\varepsilon)$ and $T_\varepsilon$ of the  ferromagnetic region, $R_A(\varepsilon = 0) = 1$: The zero-energy mode is always perfectly Andreev reflected in a hybrid structure based on helical liquids. 

Two comments are in order now. First, this result is quite different from the case of conventional N-S junctions, where backscattering can be induced also by non-ferromagnetic impurities leading to $R_A(\varepsilon=0)=T_0^2/(2-T_0)^2$\phantom{\;,}~\cite{Beenakker94}. It is worth noticing that the difference in the result originates in Eq.~(\ref{Eq:scatt_e_h}), which leads to a minus sign in front of the first square root in the denominator of Eq.~(\ref{Eq:RA-fin-gen}). The second comment is that, interestingly, the condition $R_A =1$ of perfect Andreev reflection  can in principle be satisfied by other, non-zero, energy modes as well. Indeed, the resonance condition for the ferromagnetic region to effectively become transparent is  
\begin{eqnarray}
\lefteqn{(\sqrt{R_\varepsilon}-\sqrt{R_{-\varepsilon}})^2 +}  & & \label{Eq:other_modes}\\
& &+4 \cos^2 \left[ \arccos\frac{\varepsilon}{\Delta_0} +\Phi_m^A(\varepsilon)\right]\sqrt{ R_\varepsilon  R_{-\varepsilon}} =0 \nonumber \;.
\end{eqnarray}

In order to illustrate the effect of the above general result, we  consider as a first simple example the case depicted in Fig.~\ref{Fig:Barrier}(a) of one single ferromagnetic barrier located between $x_1$ and $x_2$,   characterized by a uniform magnetization
\beq
\mathbf{m}(x) = \left\{
\begin{array}{lll}
(m_\parallel \cos \phi, m_\parallel \sin \phi, m_{z})    & \mbox{ if $x_{1} \leq x \leq x_{2}$}, \\
0                                             & \mbox{ else}.
\end{array}
\right. \nn
\eeq
In this case, the parameters of the scattering matrix (\ref{Eq:S0-p-gen}) acquire the following expressions. The phase $\Gamma_m$ reads
\begin{equation}
\Gamma_m(\varepsilon)=\arctan \left[X(\varepsilon)\right]-\frac{(\mu+\varepsilon)L_m}{\hbar v_F}\label{Gammam_sgl}
\end{equation}
with
\begin{eqnarray}
X(\varepsilon)= \left\{ \begin{array}{ll} 
\frac{(\mu+E) \tanh\left[ \frac{L_m}{\hbar v_F} \sqrt{m_{||}^2-(\mu+E)^2}\right]}{\sqrt{m_{||}^2-(\mu+E)^2}}\;,   \hspace{0.3cm} & |\mu+\varepsilon| < m_{||} \\ &  \\
\frac{(\mu+E) \tan\left[ \frac{L_m}{\hbar v_F} \sqrt{(\mu+E)^2-m_{||}^2}\right]}{\sqrt{(\mu+E)^2-m_{||}^2 }}\;, \hspace{0.3cm} & |\mu+\varepsilon| >m_{||}\;, 
\end{array}\right.  \label{Xepsilon}
\end{eqnarray}
and $L_m=x_2-x_1$ denoting the length of the barrier. The transmission coefficient reads
\begin{equation}
T_\varepsilon= \left(1+ Y_\varepsilon^2  \right)^{-1} \label{Eq:T_single_barrier}
\end{equation}
with
\begin{eqnarray}
Y_\varepsilon &=&\! \left\{ \begin{array}{ll} 
  \frac{m_{||}\sinh \left[ \frac{L_m}{\hbar v_F} \sqrt{m_{||}^2-(\mu+\varepsilon)^2}\right]}{\sqrt{m_{||}^2-(\mu+\varepsilon)^2}}\;,  \hspace{0.5cm} & |\mu+\varepsilon| < m_{||} \\ &  \\
\!   \frac{m_{||}\sin  \left[ \frac{L_m}{\hbar v_F} \sqrt{(\mu+\varepsilon)^2-m_{||}^2}\right]}{\sqrt{(\mu+\varepsilon)^2-m_{||}^2}}\;,  \hspace{0.5cm} & |\mu+\varepsilon| >m_{||} \;,
\end{array}\right. \label{Y-SGL}
\end{eqnarray}
whereas $\chi_m(\varepsilon) \equiv \chi_z=m_z L_m/\hbar v_F$, $\Phi_m(\varepsilon) = 2 x_0 (\mu+ \varepsilon)/\hbar v_F + \phi $, with $x_0=(x_1+x_2)/2$ denoting  the center of the barrier, implying $\Phi_m^A(\varepsilon) = 2\varepsilon x_0/\hbar v_F$ in Eqs.~(\ref{Eq:RA-fin-gen}) and (\ref{Eq:other_modes}). 
Notice that, at the Dirac point, $\mu = 0$, $T_\varepsilon = T_{-\varepsilon}$ and Eq.~\eqref{Eq:other_modes} reduces to 
\beq
\cos^2 \left[ \arccos\frac{\varepsilon}{\Delta_0} + \frac{2\varepsilon x_0}{\hbar v_F}\right] (1-T_\varepsilon) = 0 \;. \label{Eq:other_modes_redux}
\eeq
As $T_\varepsilon \neq 1$ for all excitation energies, the perfectly Andreev reflected modes are those for which the energy satisfies
\beq
2\arccos\frac{\varepsilon}{\Delta_0} + \frac{4\varepsilon x_0}{\hbar v_F} = \pi + 2n\pi\; \label{Eq:resonant_states}
\eeq
with $n$ an integer. The latter equation turns out to be the condition for bound-states to appear between the superconducting interface at $x=0$ and a virtual infinite ferromagnetic wall at $x=x_0$. For $\mu \neq 0$, the resonances are no longer   perfect. Nevertheless, at low energy $|\mu+E| \ll m_{\parallel}$,  the transmission coefficient becomes energy independent  and  these modes are almost perfectly Andreev reflected.


\subsection{A case study: the ferromagnetic quantum dot}

\begin{figure}
        \centering
        \subfloat{\includegraphics[width=\columnwidth,clip]{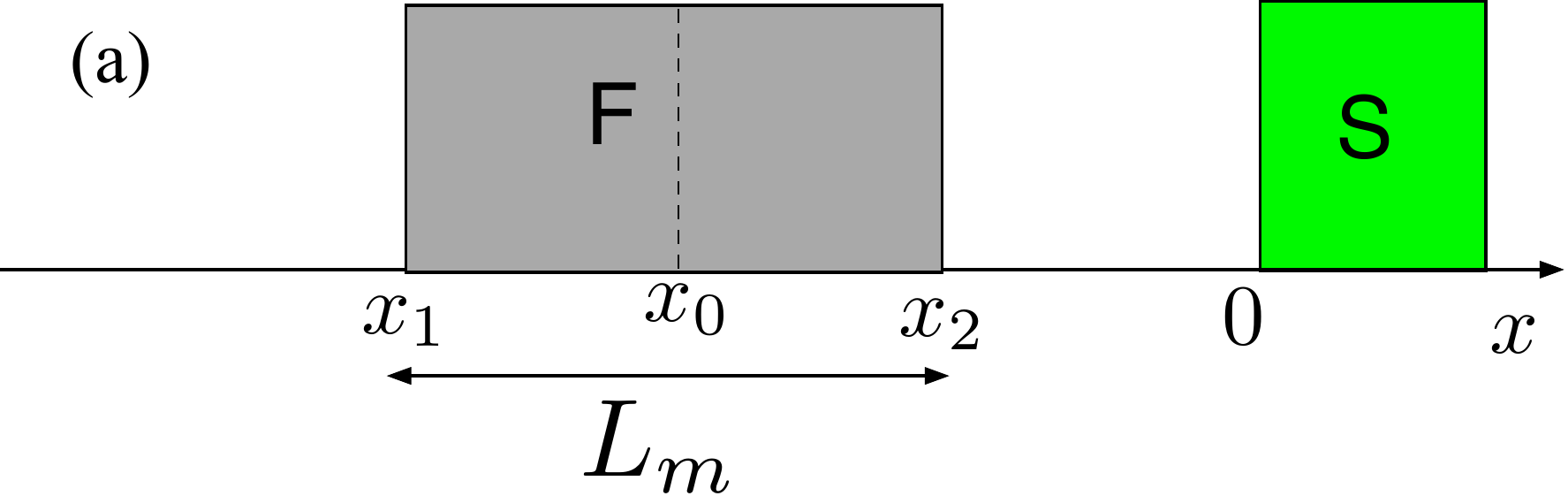}}
        
		\subfloat{\includegraphics[width=\columnwidth,clip]{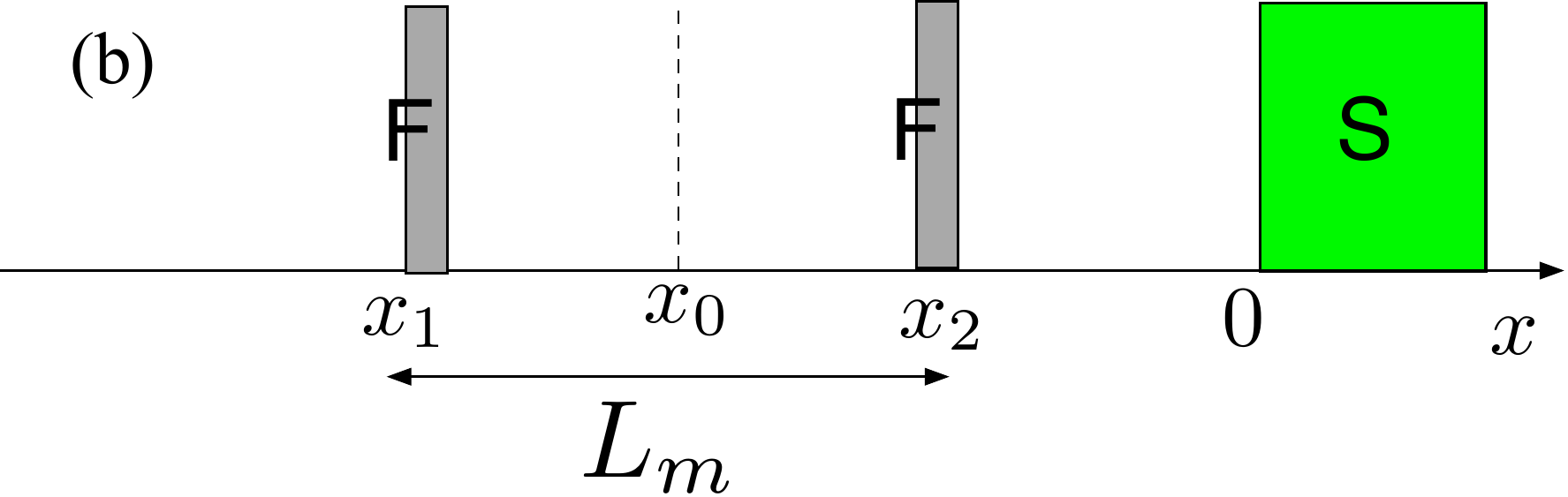}}
		 \caption{(Color online) N-S junction with (a) a ferromagnetic barrier, (b) two ferromagnetic impurities, determining a ferromagnetic quantum dot. }\label{Fig:Barrier}
\end{figure}

\begin{figure}
        \centering
        \subfloat[$\mu _0=0.5, \mu =0, L_m=1, x_0=1.5, \Delta \phi =0$]{\includegraphics[width=\columnwidth,,clip]{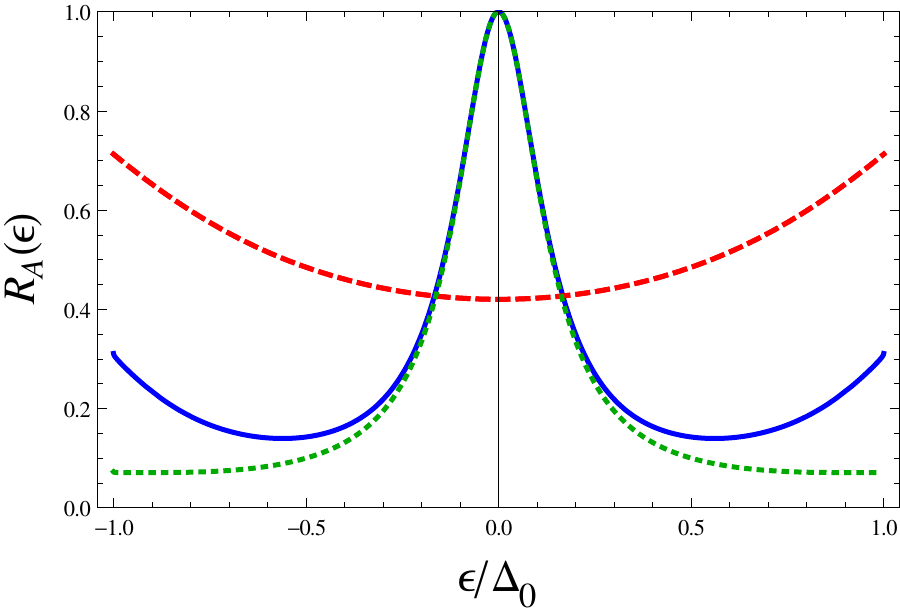}}
        
		\subfloat[$\mu _0=0.5, \mu =0, L_m=1, x_0=4.5, \Delta \phi =0$]{\includegraphics[width=\columnwidth,,clip]{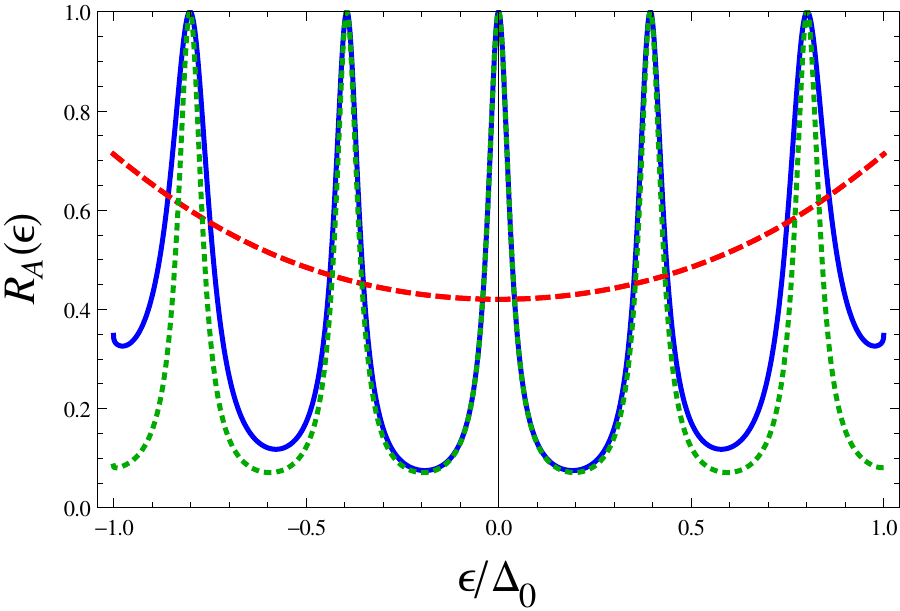}}
		 \caption{(Color online) Influence of the position $x_0$ of the ferromagnetic region on the Andreev reflection probability, in the double-barrier case (blue solid line). For a better interpretation, we also plotted the transmission probability $T_\varepsilon$ and $T_{-\varepsilon}$ of the double barrier (red dashed line) and the Andreev reflection probability for a single impurity of strength $2\mu_0$ located at $x_0$ (green dotted line). Energies are given in units of $\Delta_0$ and lengths in units of the superconductor coherence length $\hbar v_F/\Delta_0$. }\label{Fig:plot_db_1}
\end{figure}

\begin{figure}
        \centering
        \subfloat[$\mu_0=0.35, \mu=0, L_m=2.15, x_0=6$] {\includegraphics[width=\columnwidth,,clip]{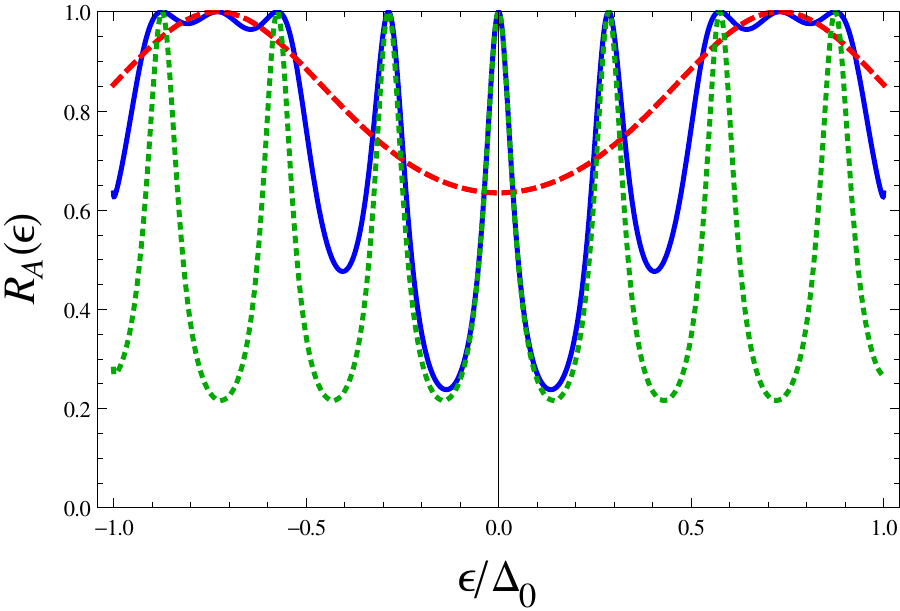}}
        
		\subfloat[$\mu_0=0.35, \mu=0.16, L_m=2.15, x_0=6$]{\includegraphics[width=\columnwidth,,clip]{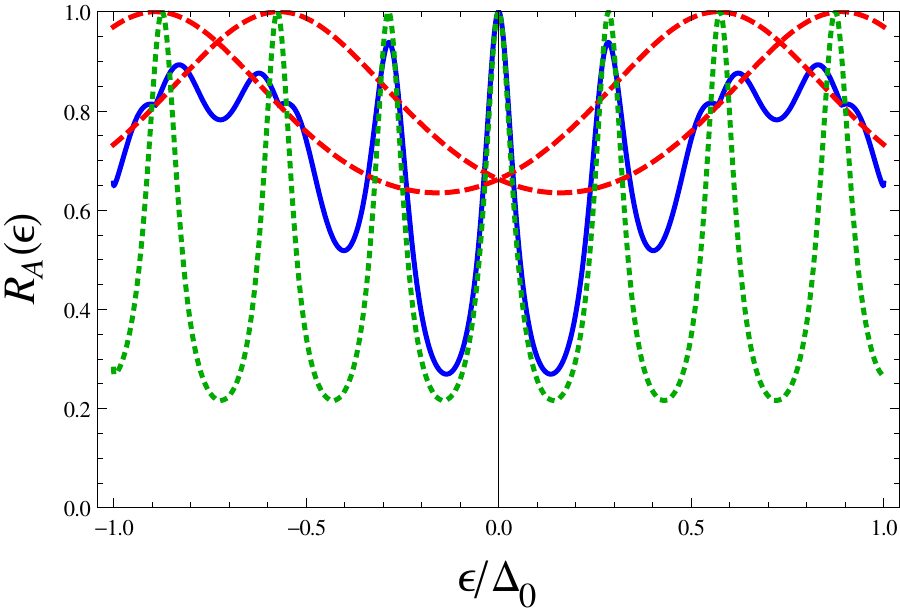}}
		
		\subfloat[$\mu_0=0.35, \mu =0.3, L_m=2.15, x_0=6$]{\includegraphics[width=\columnwidth,,clip]{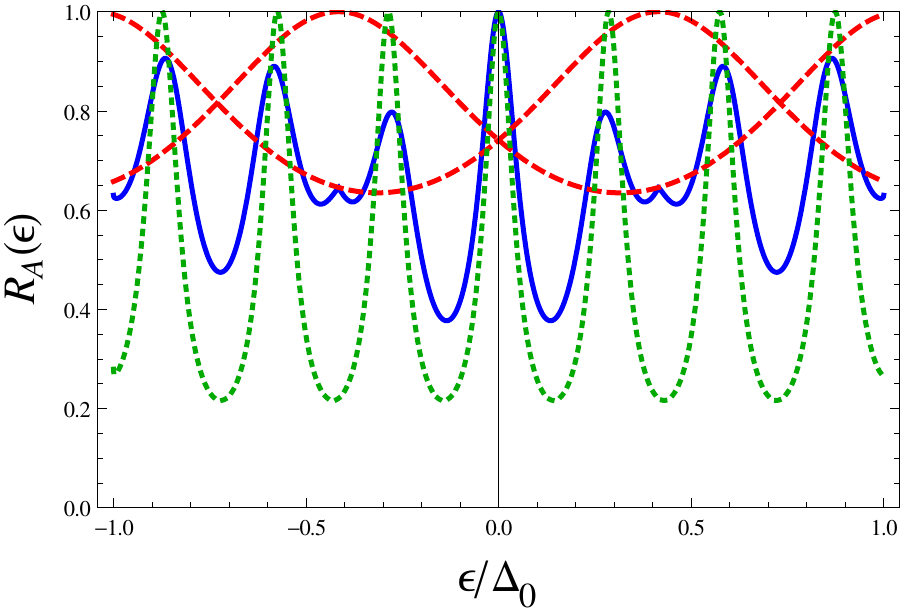}}
		
		 \caption{(Color online) Influence of the chemical potential $\mu$ on the Andreev reflection probability, in the double-barrier case (blue solid line). In all three plots, $\Delta\phi=0$. For a better interpretation, we also plotted the transmission probability $T_\varepsilon$ and $T_{-\varepsilon}$ of the double barrier (red dashed line) and the Andreev reflection probability for a single impurity of strength $2\mu_0$ located in $x_0$ (green dotted line). Energies are given in units of $\Delta_0$ and lengths in units of the superconductor coherence length $\hbar v_F/\Delta_0$. }\label{Fig:plot_db_2}
\end{figure}

As an illustration of the generality of Eq.~\eqref{Eq:RA-fin-gen}, we consider the   case where the ferromagnetic region consists of two impurities located at $x_1$ and $x_2$, described as two   barriers of  size $\delta$, as displayed in Fig.~\ref{Fig:Barrier}(b), which we call  ferromagnetic quantum dot~\cite{Sassetti13}. The center of the dot is located at $x_0 = (x_1 +x_2)/2$. The magnetic texture is then taken as follows:
\beq
\mathbf{m}(x) = \left\{
\begin{array}{lll}
(m_1 \cos \phi_1, m_1 \sin \phi_1, m_{z1})    & \mbox{ if $x_{1-} \leq x \leq x_{1+}$}, \\
(m_2 \cos \phi_2, m_2 \sin \phi_2, m_{z2})    & \mbox{ if $x_{2-} \leq x \leq x_{2+}$}, \\
0                                             & \mbox{ else}\;,
\end{array}
\right. \nn
\eeq
with $x_{1\pm} = x_1 \pm \delta/2$ and $x_{2\pm} = x_2 \pm \delta/2$. We are interested in the limit of sharp barriers, i.e. $\delta \rightarrow 0$ and $m_{1,2} \rightarrow \infty$, with keeping $\mu_1 = m_1 \delta /(\hbar v_F)$ and $\mu_2 = m_2 \delta /(\hbar v_F)$ fixed and finite. We shall restrict ourselves to the case of equal barriers, $\mu_1 = \mu_2 \equiv \mu_0$, which captures the main features of the scattering problem. Combining the scattering matrices of the two single barriers, one can straightforwardly obtain   the transmission probability of the double-barrier as
\beq
T_\varepsilon = \frac{1}{\cosh^2(2\mu_0) -\sin^2(k^{e}_\varepsilon L_m+\Delta \phi/2)   \sinh^2(2\mu_0)} \label{Eq:TE-DBL}
\eeq
with $k^{e}_\varepsilon = (\varepsilon + \mu)/\hbar v_F$, the electron wave-vector in the normal region, $L_m = x_2-x_1$ and $\Delta \phi = \phi_2 -\phi_1$. It appears that, for a fixed value of $L_m$, the chemical potential and the phase difference play a similar role, that is, breaking the symmetry between the transmission of particles and holes. In order to simplify the analysis in the various examples below, we vary only $\mu$, and take $\Delta \phi =0$. One should also note that for $L_m = 0$ and $\Delta \phi =0$,  $T_\varepsilon$ in Eq.~\eqref{Eq:TE-DBL} reduces to the transmission probability of a single impurity of strength $2\mu_0$, located at $x_0$ (compare with Eqs.~\eqref{Eq:T_single_barrier}-\eqref{Y-SGL}). For completeness we also provide here the other parameters of the quantum dot scattering matrix, namely
\begin{align}
\Gamma_m(\varepsilon) &= - \arctan \left( \frac{\sin(2k^e_\varepsilon d + \Delta \phi)\sinh^2 \mu_0}{1 + 2 \sinh^2 \mu_0 \cos^2(k^e_\varepsilon d + \Delta \phi/2) } \right)\;, \nn \\
\Phi_m(\varepsilon) &= 2 k^e_\varepsilon x_0 + \phi_0\;, \quad \phi_0 = (\phi_1+\phi_2)/2\;,\nn \\
\chi_m(\varepsilon) &= \chi_{z1} + \chi_{z2}\;, \label{Eq:param_QD}
\end{align}
with $\chi_{zi} = m_{zi}\delta/(\hbar v_F)$, that we keep finite as $\delta \rightarrow 0$. \\

In Figs.~\ref{Fig:plot_db_1} and~\ref{Fig:plot_db_2} we plot the Andreev reflection probability as a function of the excitation energy, for different values of the parameters. The various cases show that the zero-mode is always perfectly Andreev reflected, consistently with Eq.~\eqref{Eq:RA-fin-gen}. Right at the Dirac point, the modes satisfying the condition of Eq.~\eqref{Eq:resonant_states} are perfectly Andreev reflected. We distinguish two kinds of such modes. First, there are the Fabry-P\'erot-like modes, whose energy satisfies the same Eq.~(\ref{Eq:resonant_states}) as for the single barrier case, whith $x_0$ now the centre of the ferromagnetic dot. The density of such modes increases with $|x_0|$, as illustrated in Fig.~\ref{Fig:plot_db_1}~(a) and (b). Second, the modes for which $T_\varepsilon = 1$ are also perfectly Andreev reflected, a possibility that does not arise in the single barrier case. Generally speaking, as compared to a single impurity in $x_0$, $R_A(\varepsilon)$ is modulated by the transmission coefficient of the double-barrier structure. Varying the chemical potential ($\mu \neq 0$) leads to several interesting modifications. While the position of the maxima corresponding to the virtual bound-states is barely altered, their amplitude is no longer $1$ -- they are not perfectly Andreev reflected anymore -- except for the zero energy mode, which remains pinned to $1$. Furthermore the peaks corresponding to the open channels of the dot now split, as $T_\varepsilon$ and $T_{-\varepsilon}$ are no longer equal. This particular evolution as a function of the chemical potential is depicted in Fig.~\ref{Fig:plot_db_2}~(a), (b) and (c).  \\


\subsection{Majorana wave-functions }

In this section, we discuss the relation between perfect Andreev reflection and the presence of Majorana states. We come back to the somewhat simpler case of a single ferromagnetic barrier, for which the transmission probability is given in Eq.~\eqref{Eq:T_single_barrier}. As discussed before, the Andreev reflection probability can have many peaks as a function of energy, the positions and number of which depend on the location $x_0$ of the center of the barrier. However, except in the special situation $\mu = 0$, only the zero-energy mode is perfectly Andreev reflected. Such a robust peak is often interpreted in the context of topological superconductivity as the signature of tunneling into a Majorana bound-state. Here the situation is more subtle, as there is no real bound-state to begin with -- contrary to the case of a genuine spinless $p$-wave superconductor.

\begin{figure}
        \includegraphics[width=8.5cm,clip]{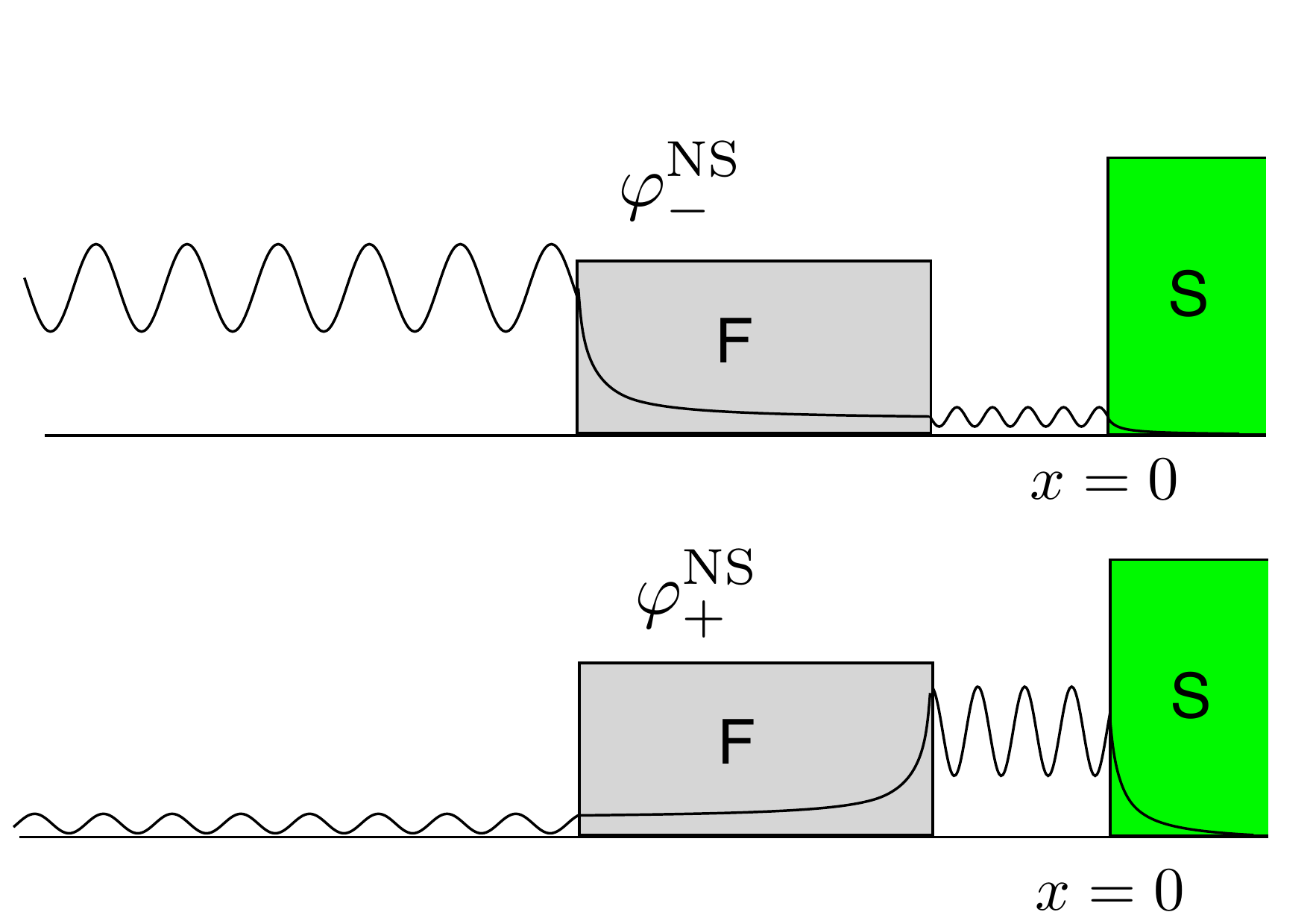}
		\caption{(Color online) Sketch of two Majorana states, at $\varepsilon = 0$, in the presence of a ferromagnetic barrier. Even  though they do extend on the whole normal region, they are predominently localized on one side or the other of the ferromagnetic domain (see text).  }\label{Fig:majorana_WF}
\end{figure}

The zero-energy states are always delocalized in the whole normal region that is ungapped. In the absence of a scattering region, Andreev reflection at the interface imposes that scattering states are superpositions of electron and hole components. The zero energy subspace is two-dimensional and spanned by two orthogonal eigenstates  of the BdG Hamiltonian, that are charge conjugated (we drop the label $\varepsilon=0$ for simplicity). The first wave function $\varphi_{1}$  corresponds to injecting a Cooper pair in the superconductor (it is the superposition of a right-moving electron and a left-moving hole). The second wavefunction $\varphi_{2}=\mathcal{C}\varphi_{1}$ will be denoted by $\varphi_{1^c}$, according to the notation of Sec.~\ref{sec:spectral}, and corresponds to the opposite process, the injection of a Cooper from the superconductor into the normal lead (it is the superposition of a left-moving electron and a right-moving hole).   Their explicit expressions read 
\beq
\varphi_{1}(x)  =  \begin{pmatrix} 1 \\ 0 \\  -i e^{-i\chi_R} \\ 0 \end{pmatrix}e^{i\mu x}\;, \quad  \varphi_{1^c}(x)  =  \begin{pmatrix} 0 \\ i e^{i\chi_R}  \\  0 \\ -1 \end{pmatrix} e^{-i\mu x}\;. \label{Eq:scatt_states}
\eeq
If one imposes a hard-wall boundary condition, somewhere in the normal region, then the only allowed solution is a superposition of $\varphi_{1}$ and $\varphi_{1^c}$ that is indeed a single Majorana state. This assumption connects the present situation to the nanowire setups where such a hard-wall boundary is usually imposed~\cite{Klinovaja12}. If one removes the hard-wall, then there are not one but two independent Majorana states, being linear combinations of $\varphi_{1}$ and $\varphi_{1^c}$. It turns out that a ferromagnetic barrier, as depicted in Fig.~\ref{Fig:Barrier}(a) can localize the two Majorana states on either side of the barrier. In order to prove this statement, we first compute the scattering states at zero energy in the presence of the barrier. Their wave-functions are given in Appendix~\ref{app:NS} and coincide in the region $x<x_1$ with the states of Eq.~\eqref{Eq:scatt_states}. Following the general scheme of Majorana states given in section~\ref{sec:spectral}, we construct two independent Majorana wave-functions, $\varphi^{\rm NS}_+$ and $\varphi^{\rm NS}_-$, given by $\varphi^{\rm NS}_\pm = \alpha_\pm \varphi_1 + \alpha_\pm^* \varphi_{1^c}$. A suitable choice of $\alpha_\pm$ --  given in Appendix~\ref{app:NS}-- leads  $\varphi^{\rm NS}_\pm$ to acquire the simple form


\begin{equation}
\varphi^{\rm NS}_\eta(x) = 
\begin{pmatrix} 
f_{\eta}(x) \, e^{\frac{i}{2} [\frac{\pi}{2}+\chi_R -\int_0^x \frac{2 m_z(x^\prime)}{\hbar v_F} dx^\prime ]} \\
f^*_{\eta}(x) \, e^{\frac{i}{2} [\frac{\pi}{2}+\chi_R -\int_0^x \frac{2 m_z(x^\prime)}{\hbar v_F} dx^\prime ]} \\
f_{\eta}(x) \, e^{-\frac{i}{2} [\frac{\pi}{2}+\chi_R -\int_0^x \frac{2 m_z(x^\prime)}{\hbar v_F} dx^\prime ]} \\
-f^*_{\eta}(x) \, e^{-\frac{i}{2} [\frac{\pi}{2}+\chi_R -\int_0^x \frac{2 m_z(x^\prime)}{\hbar v_F} dx^\prime ]}
\end{pmatrix}\;, \label{Eq:majo-N-S-compact}
\end{equation}
with $\eta=\pm$, 
\begin{equation}
f_{\eta}(x)=\frac{\, e^{-i\frac{\phi+\eta \tilde{\theta}_0}{2}}}{2 \cosh\frac{\kappa L_m}{2}} \left\{ 
\begin{array}{lcl}
e^{i k_F (x-x_1)} \, e^{-\eta  \frac{\kappa L_m}{2}} \;,     & & x \le x_1 \\
e^{\eta \kappa (x-x_0)} \;,      & &   x_1 \le x \le x_2 \\
e^{i k_F (x-x_1)} \, e^{\eta \frac{\kappa L_m}{2}} \;,    & & x \ge x_2
\end{array}\right. \label{Eq:f_eta-N-S}
\end{equation}
and $\kappa = \sqrt{m_\parallel^2 - \mu^2}/\hbar v_F $, $k_F=\mu/\hbar v_F$, $\tilde{\theta}_0 = \arccos(\mu/m_\parallel)$, $\mu < m_\parallel$. One can easily check that the $\varphi^{\rm NS}_{\eta = \pm}$ are indeed invariant under charge-conjugation, that is, $ \mathcal{C} \varphi^{\rm NS}_{\pm} = \varphi^{\rm NS}_\pm$, with $ \mathcal{C} = \mathcal{K}\; \tau_y \otimes \sigma_y$. Note that the two wave-functions have opposite exponential variations inside the ferromagnetic region $x_1 \le x \le x_2$. Although both Majorana states are extended over the whole normal region, they are still spatially localized on opposite sides of the ferromagnetic domain, as drawn schematically in Fig.~\ref{Fig:majorana_WF}.    In contrast, the zero-energy Andreev states $\varphi_1$ and $\varphi_{1^c}$ are   mixtures of these two wavefunctions and hence cannot be considered as localized in any meaningful way.  We will see in the next section that adding a second superconducting electrode barely affects these Majorana states -- they are by construction in an equal weight superposition of electron and hole and ready to be bound by a superconducting mirror. \\


\section{S-N-S (Josephson) junctions}
\label{sec:SNS}

\subsection{Andreev bound states and Josephson current}

We now turn our attention to the case of S-N-S junctions with an arbitrary ferromagnetic domain in the normal region. Aiming at drawing analogies with the former case of the N-S junction, we focus on subgap transport and start by deriving the condition for Andreev bound states~\cite{Kulik70}. We can use most of the results of the previous section. Scattering amplitudes in region   $N_2$ are still connected by the electron and hole scattering matrices as in Eq.~\eqref{Eq:scatt_amplitudes}. One only needs to implement a similar condition for perfect Andreev reflection in region $N_1$. Moreover, for the SNS case it is more convenient to set the origin at the center of the junction, so that the two interfaces are located at $x=\pm L/2$, with $L$ denoting the interface distance. The whole Andreev reflection process at both interfaces can be written as 
\begin{align}
(a_{e,1},a_{e,2},a_{h,1},a_{h,2})^T =  S_A(\varepsilon)(b_{e,1},b_{e,2},b_{h,1},b_{h,2})^T\;, \label{Eq:A_scatt_amplitudes} 
\end{align}
with 
\begin{equation}
S_A(\varepsilon)=\begin{pmatrix}  0 & \alpha^\prime(\varepsilon) \mathsf{r}_A \\  \alpha^\prime(\varepsilon) \mathsf{r}_A^* & 0  \end{pmatrix} \;
\end{equation}
\\denoting a $4\times4$ matrix where $\alpha^\prime(\varepsilon) =  \exp [-i \textrm{arccos} (\varepsilon/\Delta_0) + i(k_\varepsilon^e-k_\varepsilon^h)L]$ and $\mathsf{r}_A = \textrm{Diag}[e^{i\chi/2},e^{-i\chi/2}]$. In the case of a helical liquid with a linear spectrum, $k^{e/h}_\varepsilon=(\mu \pm E)/\hbar v_F$ and we simply have $k_\varepsilon^e-k_\varepsilon^h = 2\varepsilon/\hbar v_F$. Combining Eq.~\eqref{Eq:scatt_amplitudes} and~\eqref{Eq:A_scatt_amplitudes}, we arrive at the well-known compatibility condition~\cite{Beenakker91}
\begin{equation}
{\rm det} \left( \tau_0 \otimes \sigma_0 - S_A(\varepsilon) \, S_N(\varepsilon)  \right)=0\;,  \label{Eq:ABS}
\end{equation}
for the Andreev bound-states (ABS). In our case, the latter equation acquires the simple form
\begin{widetext}
\beq
\cos^2 \left[ \arccos\frac{\varepsilon}{\Delta_0}-\frac{\varepsilon (L-\lambda_m^S(\varepsilon))}{\hbar v_F}  \right] =\frac{1}{2} \left( 1- \sqrt{R_{\varepsilon} R_{-\varepsilon}} \, \cos\left(   2 \Phi_m^A(\varepsilon)\right)+ \sqrt{T_{\varepsilon} T_{-\varepsilon}}  \, \cos(\chi-2\chi^S_m(\varepsilon))  \right)\;, \label{Eq:ABS_2}
\eeq
\end{widetext}
where the odd function  of the energy $\Phi_m^A(\varepsilon) = (\Phi_m(\varepsilon)-\Phi_m(-\varepsilon))/2$, as well as the even functions    $\lambda_m^S(\varepsilon) =  \hbar v_F (\Gamma_m(-\varepsilon)-\Gamma_m(\varepsilon))/2 \varepsilon$ and $\chi_m^S(\varepsilon) =   (\chi_m(\varepsilon)+\chi_m(-\varepsilon))/2$  are directly extracted from the scattering matrix (\ref{Eq:S0-p-gen}) describing the ferromagnetic scatterer. Equation~\eqref{Eq:ABS_2} thus determines the Andreev bound levels in the presence of an arbitrary ferromagnetic scatterer and represents another important result of the paper.

\begin{figure*}
       \includegraphics[width=15cm,clip]{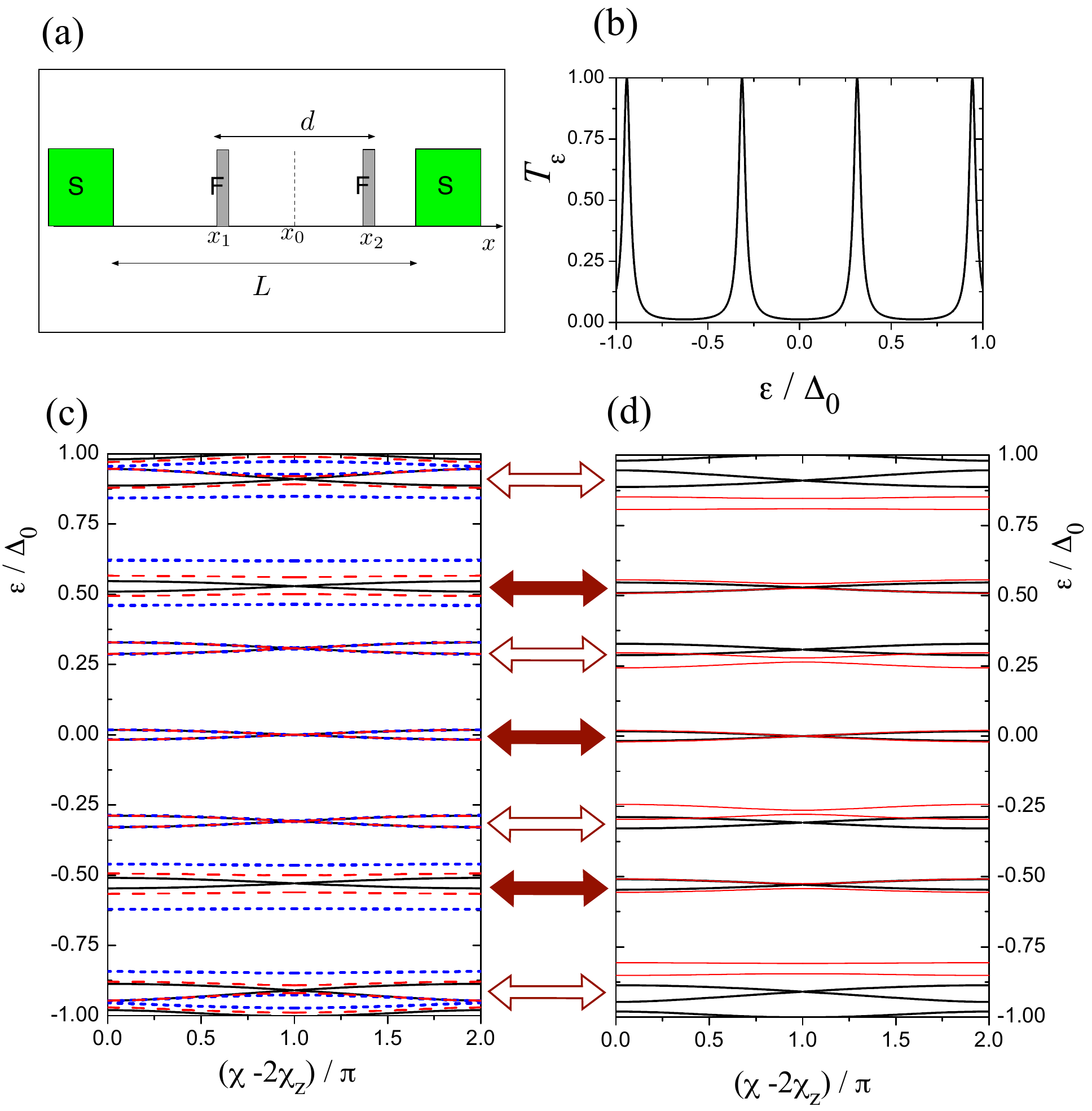}
		 \caption{(Color online) Andreev bound levels for the case of a ferromagnetic quantum dot. (a) sketch of the realisation of the system, where two ferromagnetic barriers are located inside the SNS junction of topological edge states. Here we have considered two delta-like barriers with the same strength parameter $\mu_0=1.44$ (corresponding to a transmission of 20\% for each barrier), separated by a distance $d=L/2$, and with equal magnetisations ($\phi_1=\phi_2=0$ and $m_{z1}=m_{z2}=m_z$). (b) The transmission coefficient $T_\varepsilon$ of the dot, plotted as a function of energy for $\mu=0$, shows resonances corresponding to the open channels of the quantum dot. (c) Andreev bound levels at the Dirac point ($\mu=0$) are plotted as a function of the superconductor phase difference $\chi$, in the case of a long junction ($L=10 \hbar v_F /\Delta_0$), for different values of the location of the dot center: $x_0/L=0$ (solid black), $x_0/L=0.02$ (dashed red), and $x_0/L=0.05$ (dotted blue). (d) The Andreev bound levels for the case of a centered quantum dot ($x_0=0$), for different values of chemical potential: $\mu=0$ (thick black) and $\mu=\Delta_0/2$ (thin red). Andreev levels appear not only around the resonances of the dot (denoted by empty arrows) but also in correspondence of off-resonance energies (filled arrows). Only the level around $\varepsilon=0$ is stable against variations of $x_0$ and $\mu$, and always carries a current. }\label{Fig:plot_ABS}
\end{figure*}

In order to illustrate its physical consequences, we exploit one enlightening example, namely the case of a ferromagnetic quantum dot  realised by two ferromagnetic barriers, as sketched in Fig.~\ref{Fig:plot_ABS}(a). The case of equal barriers captures the main physical ingredients of the problem and we shall restrict to this situation. The functions $T_\varepsilon$, $\Phi^A_m(\varepsilon)$, $\chi^S_m(\varepsilon)$ and $\lambda^S_m(\varepsilon)$ appearing in Eq.(\ref{Eq:ABS_2}) are in this case straightforwardly obtained from the scattering matrix of the dot, given at the end of Sec.~\ref{sec:NS-a}. In particular, Eq.(\ref{Eq:TE-DBL}) yields the transmission coefficient $T_\varepsilon$, whereas from  Eq.~\eqref{Eq:param_QD} one obtains $\Phi_m^A(\varepsilon) =2 \varepsilon x_0/\hbar v_F$, $\chi_m^S(\varepsilon) \equiv \chi_{z}$  and $\lambda^S_m(\varepsilon)$ through $\Gamma_m(\varepsilon)$.  The Andreev bound levels obtained from the solution of Eq.(\ref{Eq:ABS_2}) are plotted in Fig.~\ref{Fig:plot_ABS}(c) and (d)  as a function of the superconducting phase difference $\chi$,   for various values of the location $x_0$ of the quantum dot center and the chemical potential $\mu$, respectively. The first emerging feature is that the levels are symmetric in energy with respect to $\varepsilon=0$. This is due to the particle-hole symmetry of the BdG equations. Indeed, from the general properties of Eq.~\eqref{Eq:ABS_2} one can easily check that, because $\Phi_m^A(\varepsilon)$ is odd and $\lambda_m^S(\varepsilon)$ and $\chi_m^S(\varepsilon)$ are even, if $\varepsilon$ is a solution of Eq.~\eqref{Eq:ABS_2}, then $-\varepsilon$ is also a solution. Secondly, the plots are symmetric in the phase difference $\chi$, around the symmetry value $\chi-2\chi_z=\pi$. Indeed, from Eq.~\eqref{Eq:ABS_2} one can see that, if a bound-state exists for a given energy at a  value $\chi_1$, another one necessarily exists at $\chi_2 = 2(\pi +  \chi_m^S(\varepsilon))-\chi_1$, so that there are two bound-states in a $2\pi$ interval centered around $\chi-2\chi_m^S(\varepsilon) = \pi$. The spectrum of Andreev bound-levels is $2\pi$-periodic with the phase difference $\chi$. However, the Andreev states do not necessarily have the same periodicity, as we shall discuss below. { We notice also that the renormalization of the superconducting phase difference as $\chi\rightarrow \chi-2\chi_z$ caused by the $m_z$ magnetisation induces a $\pi$-junction behaviour when $\chi_z \gtrsim \pi/2$. The third feature emerging from Fig.~\ref{Fig:plot_ABS}  is the existence of crossing points. To discuss their physical meaning, it is worth recalling that, differently from conventional S-N-S junctions, here  for each value of $\chi-2\chi_z$ the Andreev levels are typically non-degenerate, due to the helical nature of the edge states. Crossing points, however, are an exception and correspond to degenerate eigenvalues of the BdG equations}. It is interesting to analyze whether the corresponding degenerate states hybridize or not. { In Fig.~\ref{Fig:plot_ABS}(c), we show the evolution of the ABS spectrum with the position $x_0$ of the dot center, at the Dirac point $\mu =0$. In order to understand the pattern of gap openings, one must bear in mind the resonances in the dot transmission probability $T_\varepsilon$, plotted in Fig.~\ref{Fig:plot_ABS}(b). On Figs.~\ref{Fig:plot_ABS}(c) and (d), empty and filled arrows indicate the ABS that are close to open and closed channels of the dot, respectively. These have very different behavior as $x_0$ is moved away from the center of the junction. Indeed, for open channels, $T_\varepsilon \simeq 1$ and $R_\varepsilon \simeq 0$, such that the condition for ABS barely depends on $x_0$, as one can see in Eq.~\eqref{Eq:ABS_2}. On the other hand, for closed channels, $T_\varepsilon \simeq 0$ and $R_\varepsilon \simeq 1$, and the condition for ABS barely depends on the phase difference anymore -- which explains the flatness of the bands. Note that, although the zero mode is in principle a closed channel of the dot, it is unaffected by changes in the position. Tuning the chemical potential away from $\mu=0$ also has the effect of opening gaps for all ABS, as can be seen in Fig.~\ref{Fig:plot_ABS}(d). Again,} the crossing point at $\varepsilon=0$ is stably preserved. The crossing point at $\varepsilon=0$ is thus the only one that is stable to any parameter variation.
This is in fact a general feature that stems from Eq.~\eqref{Eq:ABS_2}, from which one can see that $\varepsilon=0$ and $\chi-2\chi^S_m(0)=\pi$ is always a solution of the ABS equation,  in sharp contrast with conventional $s$-wave junctions, where normal backscattering opens a gap at zero energy. The crossing at zero energy is protected because the two Andreev states, being charge-conjugated to each other, have different fermion parity. Indeed, following Eq.~(\ref{H-equiv-2}), the Hamiltonian for the two states crossing zero energy can be written as
\beq
H_{\textrm{ABS},0} = \frac{1}{2} \varepsilon_0(\chi) \Gamma_0^\dagger \Gamma_0^{} - \frac{1}{2} \varepsilon_0(\chi) \Gamma_{0^c}^\dagger \Gamma_{0^c}^{}\;, 
\eeq
or, using $\Gamma_{0^c} = \Gamma_0^\dagger$ following from particle-hole symmetry~\cite{Fu09b}, 
\beq
H_{\textrm{ABS},0} =  \varepsilon_0(\chi) \left( \Gamma_0^\dagger \Gamma_0^{} - \frac{1}{2} \right) + \frac{1}2{}\varepsilon_0(\chi)  \;, 
\eeq
similar to Eq.~(\ref{H-equiv-1}). The two Andreev states, with energy $\pm \varepsilon_0(\chi)$ correspond to the two parity sectors, $\Gamma_0^\dagger \Gamma_0^{} = 0,1$. Such a protection directly affects the Josephson current. Indeed, Andreev bound-states carry a stationary supercurrent accross the junction, as the two Andreev reflections have the effect of transferring a Cooper pair from one superconducting contact to the other one. At zero temperature, each ABS contributes $J_n = (e/h)   \partial_\chi \varepsilon_n(\chi)$ to the total Josephson current. Levels with opposite energies therefore carry opposite supercurrents, as do degenerate levels on opposite sides of $\chi-2\chi_m^S(\varepsilon) = \pi$.   As a consequence of the protected crossing at the $\varepsilon=0$ level, although the spectrum is $2\pi$ periodic, the Josephson current is only $4\pi$ periodic. Indeed, while higher energy Andreev levels contribute a $2\pi$ periodic Josephson current, the current carried by this level is actually $4\pi$ periodic, a signature of the fermion parity anomaly in helical Josephson junctions~\cite{Kitaev01, Fu09b, Crepin13b}. { Note that the $2\pi$ current can be considerably reduced, almost filtered out, by the presence of an off-centered quantum dot, as many high energy levels become flat.}







We conclude this section by observing that  for the case of a single barrier  analytic expressions for the ABS can be determined 
in the limit of strong in-plane magnetization $m_\parallel \gg \mu, |\varepsilon|$. Indeed in this limit the transmission probability $T_\varepsilon$ of the single barrier  [see Eq.~\eqref{Eq:T_single_barrier}]  becomes energy-independent and reduces to $T_\varepsilon \rightarrow T_\infty = 1/\cosh^2 \mu_0$, with $\mu_0 = m_\parallel L_m/\hbar v_F$ parametrizing the strength of the barrier, whereas the length scale $\lambda_m^S$  reduces to $\lambda_m^S(\varepsilon) \rightarrow   L_m$.   
In the special case of a  barrier  centered in the middle of the junction, $x_0 = 0$, and at $\chi -2\chi_z = \pi$, the position of Andreev bound states is simply given by 
\beq
\frac{\varepsilon (L-L_m)}{\hbar v_F} -\arccos\frac{\varepsilon}{\Delta_0}= -\frac{\pi}{2} +m \, \pi\;, \label{ABS-special}
\eeq 
which is the analogue of the condition \eqref{Eq:resonant_states} for resonant states in the N-S junction. In this limit the positions of all these ABS (not just the one at $\varepsilon=0$) are insensitive to the strength $\mu_0$ of the barrier. When $L_m = L$ we recover the limit studied by Fu and Kane in Ref.~\onlinecite{Fu09b} and only the zero energy mode is pinned. The other extreme limit of $L_m =0$ corresponds to the impurity studied in Ref.~\onlinecite{Zhang13}. Interestingly, the relevant length scale in the problem is $L-L_m$. Eq.~(\ref{ABS-special}) shows that the condition for the definition of short and long junctions should actually be formulated in terms of the interface length $L-L_m$. The short junction limit would correspond to $L - L_m \ll \hbar v_F/\Delta_0$, while the long junction would correspond to  $L - L_m \gg \hbar v_F/\Delta_0$. In particular, the density of Andreev bound-states will be set by the length $L-L_m$.   
In the   short junction limit, one can also show that there are only two Andreev bound levels, given by 
\begin{equation}
\varepsilon_\pm(\chi) = \pm \Delta_0 \sqrt{T_0} \cos(\frac{\chi}{2}-\chi_z) 
\end{equation}
which show  the $4\pi$-periodicity, in agreement with the results by Fu \& Kane~\cite{Fu09b} and Kwon {\it et al.}~\cite{Kwon04}. This result should be compared with the short-junction limit for conventional S-N-S junctions, $\varepsilon_\pm(\chi)=\pm \Delta_0  (1-T_0 \sin^2(\chi/2))^{1/2}$, which is $2\pi$-periodic~\cite{Beenakker91}. It is worth emphasising that such a difference in the results stems from the minus sign in front of the $\sqrt{R_\varepsilon R_{-\varepsilon}}$ on the right hand side of Eq.~(\ref{Eq:ABS_2}). Similarly to the case of the N-S junction, this sign is a consequence of  Eq.~(\ref{Eq:scatt_e_h})\;.

\subsection{Majorana wave-functions}

\begin{figure}
        \includegraphics[width=8.5cm,clip]{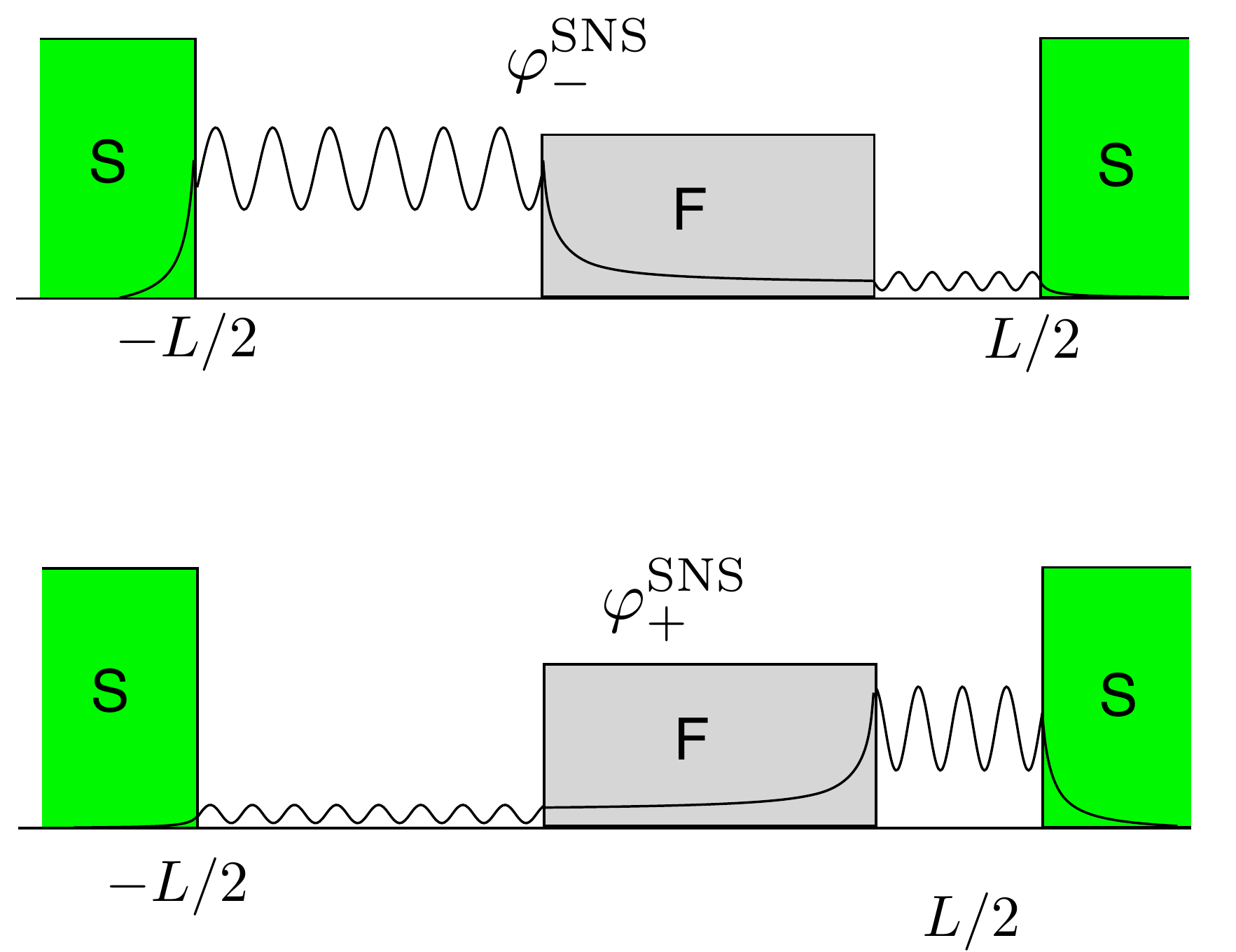}
		\caption{(Color online) Sketch of the two Majorana states at $\varepsilon = 0$ and $\chi - 2\chi_s(0) = \pi$. Similar to the NS case (see Fig.~\ref{Fig:majorana_WF}), even though they do extend on the whole normal region, they are predominently localized on one side or the other of the ferromagnetic domain. }\label{Fig:majorana_WF_2}
\end{figure}

We close our analysis with a discussion of the Majorana wave-functions in the S-N-S case. A comparison with the N-S junction is quite enlightening here. We know from the latter case that, at zero energy, there are two charge-conjugated Andreev states corresponding to a right-moving electron being reflected as a left-moving hole and a right-moving hole being reflected as a left-moving electron. The extra superconducting electrode transforms these two extended Andreev states into Andreev bound states, carrying opposite supercurrents. Again, one can decompose this single, zero energy fermionic level into two Majorana wave-functions $\varphi_+$ and $\varphi_-$ given by
\begin{equation}
\varphi^{\rm SNS}_{\eta=\pm}(x) = 
\left( 
\begin{array}{lcl} 
f_{\eta}(x) \, e^{-i \int_{x_0}^x \frac{m_z(x^\prime)}{\hbar v_F} dx^\prime  } \\
f^*_{\eta}(x) \, e^{-i \int_{x_0}^x \frac{m_z(x^\prime)}{\hbar v_F} dx^\prime  }  \\
f_{\eta}(x) \, e^{+i \int_{x_0}^x \frac{m_z(x^\prime)}{\hbar v_F} dx^\prime  }  \\
-f^*_{\eta}(x) \, e^{+i \int_{x_0}^x \frac{m_z(x^\prime)}{\hbar v_F} dx^\prime  } 
\end{array}
\right)\;, \label{Eq:majo-S-N-S-compact}
\end{equation}
with
\begin{equation}
f_{\eta}(x)=\frac{\, e^{-i\frac{\phi+\eta \tilde{\theta}_0}{2}}}{2 \cosh\frac{\kappa L_m}{2}} \left\{ 
\begin{array}{lcl}
e^{i k_F (x-x_1)} \, e^{-\eta  \frac{\kappa L_m}{2}} \,    & & x \le x_1\;, \\
e^{\eta \kappa (x-x_0)} \,    \,  & &   x_1 \le x \le x_2\;, \\
e^{i k_F (x-x_2)} \, e^{\eta \frac{\kappa L_m}{2}} \,    & & x \ge x_2\;,
\end{array}\right. \label{Eq:f_eta-S-N-S}
\end{equation}
and $\kappa = \sqrt{m_\parallel^2 - \mu^2}/\hbar v_F $, $\tilde{\theta}_0 = \arccos(\mu/m_\parallel)$, $\mu < m_\parallel$. Notice that  the wavefunction $f_\eta(x)$, which depends on the {\it in-plane} magnetization $m_{\parallel}$,  is the {\it same} for the S-N-S case, Eq.~\eqref{Eq:f_eta-S-N-S}, and for the N-S case, Eq.~\eqref{Eq:f_eta-N-S}. The difference between $\varphi^{\rm NS}_{\eta}(x)$ in Eq.~\eqref{Eq:majo-N-S-compact} and $\varphi^{\rm SNS}_{\eta}(x)$ in Eq.~\eqref{Eq:majo-S-N-S-compact}  lies in the other phase factors, that arise from the phase difference accross the junction and the $m_z$ magnetization only. As in the NS case, although extended in the whole normal region these wave-functions are localized on opposite sides of the ferromagnetic domain (see Fig.~\ref{Fig:majorana_WF_2} for a schematic illustration). The fact that two Majorana states arise in such a S-N-S junction can be contrasted with a similar situation in topological nanowire junctions. There again, two Majorana bound states exist on their own at the edges of the superconductors. When a junction is formed, they simply delocalize in the whole normal region~\cite{Chevallier12}. In the present case, helicity combined with fermion parity conservation, protects the zero energy crossing and allow for the appearance of Majorana states, that can be localized by a ferromagnetic domain. What is more surprising is that one superconducting contact alone is able to {\it preform} such localized states. 


\section{Conclusion}
\label{sec:conclu}

We have studied transport properties of hybrid structures based on helical liquids at the edge of a quantum spin Hall insulator. We explicitly computed the Andreev reflection coefficient for N-S junctions and the condition for Andreev bound states in S-N-S junctions, in both cases in the presence of an arbitrary ferromagnetic scatterer. We found that many peaks, and not only a zero-bias peak,   arise in the conductance measurement of N-S junctions, due to Fabry-P\'erot like resonances. The height of these peaks depends on external, possibly controllable, parameters, like the chemical potential or the form of the ferromagnetic barrier. In particular, the response of the double barrier setup, that we studied in detail, is very sensitive to the value of the chemical potential, which can in principle be controlled by an external gate. As the gate is varied, while some peaks   change positions and height and others even split, the zero-bias peak remains pinned. This effect should provide an experimental test to probe the peculiar and very rich interplay of helicity and superconductivity at the edge of a topological insulator, as well as to single out evidence of the Majorana zero modes. We have also shown, by computing the wave-functions, {that the presence of a ferromagnetic domain  already localizes two Majorana modes at the N-S interface. Adding a second superconducting contact binds them in a finite size S-N-S Josephson junction.} There, the two Majorana states hybridize, forming an Andreev level. We have also analyzed the general structure of the Andreev bound states spectrum, for an arbitrary ferromagnetic region. We found that the effective phase difference across the junction as well as the effective length of the junction are renormalized in an energy-dependent way by the scatterer, the latter leading to a redefinition of the short and long junction limits in the strong barrier case. Degenerate levels, manifested as crossing points in the spectrum at a phase difference of $\pi$, appear in the case of a barrier exactly centered in the junction. However, only the zero-energy crossing is truly protected due to fermion parity conservation, and as a consequence, the Josephson current across the junction is $4\pi$ periodic, a hallmark of the edge states helicity.


\section*{ACKNOWLEDGMENTS}

We would like to acknowledge financial support by the DFG (German-Japanese research unit "Topotronics" and SPP 1666) as well as the Helmoltz Foundation (VITI). {F.D. thanks the University of W\"urzburg for financial support and  hospitality during his stay as a guest professor, and also acknowledges FIRB 2012 project HybridNanoDev (Grant No.RBFR1236VV).} 
\onecolumngrid
\appendix

\section{Wave-functions in the zero-energy subspace of NS junctions}
\label{app:NS}

We present here the wave-functions of the two zero-energy scattering states, in the case of the N-S junction with a single ferromagnetic barrier, as shown in Fig.~\ref{Fig:Barrier}(a). Since the zero energy modes are perfectly Andreev reflected, in region $x<x_1$ the wave-functions are either superpositions of an incoming electron and a reflected hole or an incoming electron and a reflected hole. The first four-component wave function, which we denote by $\varphi_{1}(x)$, corresponds to the injection of a Cooper pair in the superconductor and has the following wave-function

{\small 
\begin{equation}
\varphi_1(x)=\left(
\begin{array}{c} u_{\uparrow,1}(x) \\ u_{\downarrow,1}(x) \\ v_{\downarrow,1}(x) \\ v_{\uparrow,1}(x) \end{array} 
\right)  =  
\left\{ 
\begin{array}{ll}
 \left(
\begin{array}{c} 1 \\ 0  \\  -i e^{-i(2 \chi_z+\chi_R)}  \\ 0 
\end{array} 
\right)  \, e^{i k_F x}  & \hspace{0.5cm} x<x_1\;, \\ & \\
\left(
\begin{array}{l}  \,\, \frac{e^{-i \frac{m_z (x-x_0)}{\hbar v_F}} e^{-i\frac{\chi_z}{2}}   }{2 \sin\tilde{\theta}_0}  i e^{i k_F x_1}  \left(  e^{\kappa (x-x_0)} e^{\frac{\kappa L_m}{2}} e^{-i\tilde{\theta}_0} -  e^{-\kappa (x-x_0)}e^{-\frac{\kappa L_m}{2}} e^{i\tilde{\theta}_0}   \right)   \\  \\ \\
\, \, e^{i\phi}\frac{e^{-i \frac{m_z (x-x_0)}{\hbar v_F}} e^{-i\frac{\chi_z}{2}}   }{2 \sin\tilde{\theta}_0}  i e^{i k_F x_1}  \left(  e^{\kappa (x-x_0)} e^{\frac{\kappa L_m}{2}} -  e^{-\kappa (x-x_0)}e^{-\frac{\kappa L_m}{2}}  \right)   \\  \\
 \\
-i e^{-i\chi_R} e^{-i\frac{2 m_z (x_2-x)}{\hbar v_F}} \, u_{0 \uparrow} \\ \\
 i e^{-i\chi_R} e^{-i\frac{2 m_z (x_2-x)}{\hbar v_F}} \, u_{0\downarrow}
\end{array} 
\right)     &   x_1<x<x_2\;, \\ & \\
 \left(
\begin{array}{c} \frac{e^{i \Gamma_0} e^{-i \chi_z}}{\sqrt{T_0}} e^{i k_F x}  \\ i   e^{-i k_F x}  \sqrt{\frac{1-T_0}{T_0}}  e^{i (2k_F x_0 +\phi)} e^{-i \chi_z} \, \\ -i e^{-i\chi_R} u_{0\uparrow} \\ i e^{-i\chi_R} u_{0\downarrow} \end{array} 
\right)     \,   
  & \hspace{0.5cm} x>x_2\;,
\end{array}
\right.
\end{equation}
}
where $\kappa = \sqrt{m_\parallel^2 -\mu^2}/(\hbar v_F)$, $\tilde{\theta}_0 = \arccos(\mu/m_\parallel)$, $T_0 = 
\left(1+  \frac{\sinh^2 \left[ \kappa  L_m \right]}{\sin^2\tilde{\theta}_0  }      \right)^{-1} $, $\Gamma_0 =    
\arctan \left( \frac{1}{\tan\tilde{\theta}_0  } \tanh\left[ \kappa  L_m \right] \right) \,- k_F  L_m$, $\chi_z = m_z L_m/(\hbar v_F)$, with $L_m = x_2 -x_1$ and $x_0 = (x_1+x_2)/2$. The second state corresponds to the reverse process of injecting a Cooper pair from the superconductor, into the normal region. We call this state $\varphi_{1^c}$ and it is simply given by $\varphi_{1^c} = \mathcal{C} \varphi_{1}$, with $\mathcal{C} = \mathcal{K} \tau_y \otimes \sigma_y$ the charge conjugation operator. From these two charge conjugated scattering states one can construct two arbitrary independent Majorana wavefunctions of the form $\varphi_\pm = \alpha_\pm \varphi_{1}  + \alpha_\pm^* \varphi_{1^c} $. A suitable choice of $\alpha_\pm$ leads to two Majorana states localized on either side of the ferromagnetic domain. We found
\beq
\alpha_\eta = \frac{e^{-i (k_F x_1-\chi_z-\chi_R/2)}e^{-i \eta \tilde{\theta_0}/2} e^{- \eta \kappa L_m/2}}{2 \cosh(\kappa L_m/2)}\;, \quad \eta = \pm \;.
\eeq

\twocolumngrid
\bibliography{Top_ins_wurzburg}

\end{document}